\documentclass{article}

\usepackage{arxiv}

\usepackage[utf8]{inputenc}
\usepackage[T1]{fontenc}
\usepackage{hyperref}
\usepackage{url}
\usepackage{booktabs}
\usepackage{multirow}
\usepackage{tabularx}
\usepackage{amsfonts}
\usepackage{amsmath}
\usepackage{nicefrac}
\usepackage{microtype}
\usepackage{cleveref}
\usepackage{graphicx}
\usepackage{natbib}
\usepackage{doi}
\usepackage{array}
\usepackage{tabularx}
\usepackage{makecell}
\usepackage[table]{xcolor}

\title{Teffic-Audio: Tell Fact from Fiction}

\date{}

\author{Amphion Team}

\renewcommand{\headeright}{Technical Report}
\renewcommand{\undertitle}{Technical Report}
\renewcommand{\shorttitle}{Teffic-Audio: Tell Fact from Fiction}

\hypersetup{
 pdftitle={Teffic-Audio: Tell Fact from Fiction},
 pdfsubject={},
 pdfauthor={Amphion Team},
 pdfkeywords={},
}

\begin{document}
\maketitle

\begin{abstract}

Speech deepfake detection has expanded in scope with increasingly heterogeneous spoofing mechanisms, including speech synthesis, voice conversion, vocoder reconstruction, and neural-codec resynthesis. The resulting spoofing artifacts can be further shaped by variability in source speech, recording environments, and transmission channels. 
This variability makes robust generalization across heterogeneous conditions a central requirement for practical detection systems.
This report presents Teffic-Audio, a general speech deepfake detection system designed for comprehensive evaluation environment. Teffic-Audio adopts a straightforward detector architecture consisting of a Conformer-based speech encoder, multi-head attentive statistics pooling, and a binary classifier. Rather than relying on additional architectural complexity, the system improves generalization through its training recipe, which integrates multi-source data, attack- and source-balanced sampling, and diverse audio augmentation.
Trained only with open-source data, Teffic-Audio achieves a pooled EER of 1.454\% on the 14 test sets of Speech-DF-Arena, outperforming all currently public systems on the leaderboard.
It also obtains the lowest EER on five individual test sets and shows a favorable performance--complexity trade-off compared with larger leading systems.
Overall, Teffic-Audio provides a strong and practical reference system for general speech deepfake detection.

\end{abstract}


\section{Introduction}

Recent advances in speech generation~\cite{ju2024naturalspeech, du2024cosyvoice, chen2024vall,chen2025f5} continue to broaden the scope of speech deepfake detection. The sources of spoofed speech are no longer confined to a small set of speech synthesis or voice conversion methods~\cite{ASVspoof2015, wang2020asvspoof2019}. They now cover generation systems that differ in development stage, technical family, and processing pipeline~\cite{muller2024mlaad,li2025survey}. Recent studies also treat vocoder-reconstructed audio~\cite{sun2023ai,frank2021wavefake} and audio resynthesized after neural codec compression and decoding~\cite{xie2025codecfake} as targets in spoofing detection or spoofing-related evaluation, which further expands the boundary of the detection problem. The artifacts produced by different generation mechanisms are often not consistent~\cite{muller2024harder}. Even when spoofing traces appear similar, their observable forms can vary with the source of bonafide speech, recording conditions, channel transmission, and compression codecs~\cite{chandra2025deepfake, shi2025benchmarking}. The central goal of speech deepfake detection therefore shifts from identifying specific spoofing methods to maintaining stable and general discriminative ability across diverse bonafide speech sources, spoofing mechanisms, and propagation conditions~\cite{muller2024harder,dowerah2026speech}.

This change in task scope is also reflected in the evolution of evaluation benchmarks, which move from specific anti-spoofing scenarios toward comprehensive generalization assessment. The early ASVspoof series~\cite{ASVspoof2015,wang2020asvspoof2019,yamagishi2021asvspoof} mainly formulates evaluation tasks around typical attack scenarios such as logical access and physical access. As speech spoofing techniques and application environments continue to change, the ADD challenges~\cite{yi2022add,yi2023add} and independent evaluation sets such as In-the-Wild~\cite{muller2022does} extend the evaluation coverage to more complex audio conditions, including low-quality speech and real-world noise. ASVspoof 5~\cite{wang2024asvspoof5} further organizes evaluation over larger-scale speech sources, attack types, and adversarial conditions, imposing stronger requirements on the generalization and robustness of detectors. Nevertheless, these evaluations remain largely tied to their own data construction strategies and protocol designs. In contrast to benchmarks centered on a single evaluation source, Speech-DF-Arena~\cite{dowerah2026speech}\footnote{\url{https://huggingface.co/spaces/Speech-Arena-2025/Speech-DF-Arena}} integrates multiple representative test sets under a unified protocol and metric system. This design makes cross-dataset stability directly comparable and provides a more suitable comprehensive environment for evaluating general speech deepfake detectors.

Existing studies improve speech deepfake detection systems along several axes to enhance robustness and generalization. Earlier methods mainly strengthen the ability of detectors to capture spoofing cues through spectral feature modeling~\cite{wu2020light,li2021replay}, end-to-end waveform modeling~\cite{tak2021end}, or spectro-temporal relation modeling~\cite{jung2022aasist}. With the development of self-supervised speech models, pretrained encoders such as wav2vec 2.0~\cite{baevski2020wav2vec}, WavLM~\cite{chen2022wavlm}, and XLS-R~\cite{babu2021xls} have become important bases for strong detectors because they provide richer acoustic and speech representations for downstream detection. Building on these encoders, recent methods further exploit discriminative information across different representation levels, temporal scales, and acoustic conditions through multi-layer feature fusion~\cite{zhang2024audio,wang2025mixture}, attentive aggregation~\cite{truong2024temporal,li2025frame}, sequence modeling~\cite{tran2025leveraging}, or expert routing~\cite{negroni2025leveraging,pan2025molex}.

Beyond model and representation design, data factors also receive increasing attention. Augmentation methods~\cite{cohen2022study} such as RawBoost~\cite{tak2022rawboost} aim to improve robustness to changes in propagation conditions by simulating channel, codec, and nonlinear distortions. Ge et al.~\cite{ge2025post} emphasize the importance of large-scale datasets for adapting pretrained SSL representations. Data-centric methods such as DOSS~\cite{huang2025data} further discuss the influence of generator diversity and heterogeneous data mixing on cross-dataset generalization. These studies demonstrate that detection performance is shaped by multiple factors. Nevertheless, existing methods are either mainly validated on specific benchmarks or limited cross-domain settings~\cite{yang2025generalizable,huang2025generalizable,huang2026generalizable}, or have not yet shown sufficiently strong overall performance in a comprehensive multi-dataset evaluation environment~\cite{kulkarni2026compactsslbackbonesmatter,laakkonen2026generalizable}. As a result, there is still limited evidence on how to build a general speech deepfake detector that can adapt to heterogeneous evaluation scenarios.

Under this background, this report introduces Teffic-Audio\footnote{Demo page: \url{https://tefficlabs.com/teffic-audio}}, a general speech deepfake detection system designed for the comprehensive evaluation environment. The system adopts a standard detector architecture, consisting of a Conformer-based~\cite{gulati2020conformer} speech encoder, multi-head attentive statistics pooling~\cite{okabe2018attentive, india2019self}, and a binary classifier. On this basis, the system focuses on the design of the training procedure, including multi-source data construction, attack- and source-balanced sampling, and diverse audio augmentation. These designs jointly contribute to constructing a more effective training distribution. Our system is trained only with open-source data and achieves a pooled EER of 1.454\% on the 14 test sets of Speech-DF-Arena; when compared with all currently public systems, this result ranks first. Meanwhile, Teffic-Audio achieves the current lowest EER on 5 individual test sets and shows a favorable performance--complexity trade-off when compared with leading systems on the current public leaderboard. These results provide direct evidence for building a general speech deepfake detector: a simple detector architecture can achieve strong generalization when supported by a well-designed training distribution, without necessarily relying on increasingly complex detection architectures.
Further ablation studies show that diverse audio augmentation is a key factor in improving the system's performance in complex and difficult scenarios. In addition, the choices of encoder backbone and pooling layer both have a significant impact on the final system performance. Notably, we also find that a system variant with only 4 Conformer blocks can still achieve a pooled EER of 3.346\%, indicating that the system can maintain strong cross-dataset generalization with substantially reduced encoder depth.

\begin{figure*}[t]
    \centering
    \includegraphics[width=\textwidth]{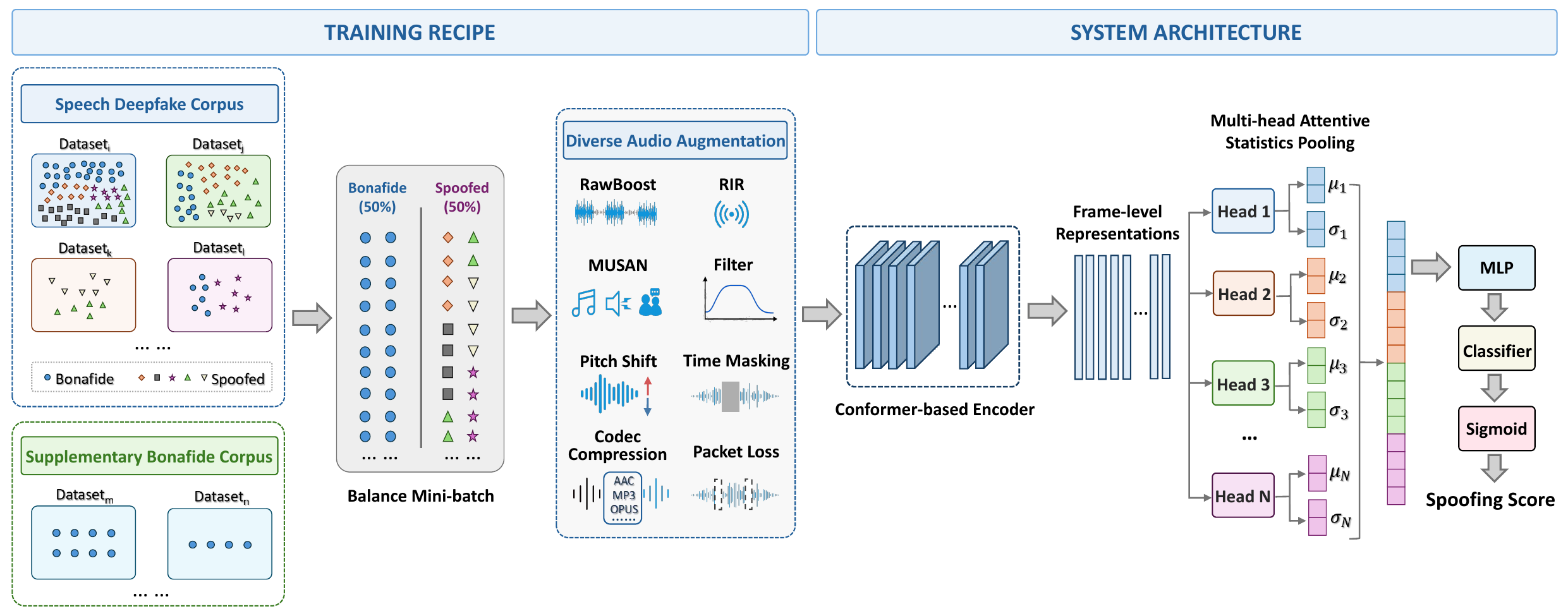}
    \caption{
    Overview of the Teffic-Audio system architecture and the  waveform-to-score training pipeline.
    }
    \label{fig:system_overview}
\end{figure*}

\section{System Overview}
\label{sec:system_overview}
The system targets utterance-level speech deepfake detection. Given an input speech utterance, the model outputs a detection score that represents the likelihood of the utterance being spoofed speech. The overall architecture consists of a SSL-initialized speech encoder, a pooling layer, and a binary classifier, forming an end-to-end discriminative model. The system architecture is illustrated on the right side of Figure~\ref{fig:system_overview}.

The speech encoder is initialized from w2v-BERT 2.0~\cite{barrault2023seamless}. Its backbone comprises a CNN-based feature extractor followed by 24 Conformer blocks~\cite{gulati2020conformer}, each with a model dimension of 1024. The encoder converts the input waveform into contextualized frame-level speech representations.

On top of the frame-level representations, the system adopts multi-head attentive statistics pooling (MHASP)~\cite{okabe2018attentive, india2019self} to obtain a fixed-dimensional utterance-level representation. MHASP uses 4 attention heads to compute attention weights along the temporal dimension and estimates the weighted mean and weighted standard deviation accordingly. The statistics from all heads are concatenated to form a 2048-dimensional pooling representation, which is then projected by an MLP with hidden dimensions of $2048 \rightarrow 1536 \rightarrow 1024$. This design aggregates discriminative information along the temporal dimension from multiple attention views.

Finally, the system uses an MLP classifier with a hidden dimension of 512 to map the 1024-dimensional global representation to binary classification logits. After sigmoid activation, the posterior probability of spoofed speech is used as the detection score. During training, the model is optimized with the binary cross-entropy (BCE) loss:
\begin{equation}
  \mathcal{L}_{\mathrm{BCE}} = -\frac{1}{N}\sum_{i=1}^{N}
  \left[y_i\log p_i + (1-y_i)\log(1-p_i)\right],
\end{equation}
where $y_i \in {0,1}$ denotes the binary label of the $i$-th sample, and $p_i$ denotes the posterior probability that the sample is predicted as spoofed speech. This objective is applied directly to the complete detection path, enabling the speech encoder, pooling layer, and classifier to be jointly optimized in an end-to-end manner. During inference, this probability is directly used as the final score without additional post-processing.

\section{Training Recipe}

The training recipe is a central component of the system design. In a comprehensive evaluation environment, the main challenge arises from distributional heterogeneity caused jointly by different spoofing mechanisms, bonafide speech conditions, and propagation chains. This section describes how the training procedure is designed to obtain a more comprehensive supervision distribution.
As illustrated on the left side of Figure~\ref{fig:system_overview}, the system is developed from three aspects: Multi-source Training Corpus extends the source coverage of bonafide and spoofed speech, Attack- and Source-Balanced Data Sampling adjusts the gradient contributions of different spoofing mechanisms and bonafide speech sources, and Diverse Audio Augmentation increases the observable variation of training samples under different acoustic environments and transmission conditions.

\subsection{Multi-source Training Corpus}
A single dataset usually reflects only limited speech sources, collection conditions, and spoofing methods, making it insufficient to support cross-domain generalization in comprehensive evaluation. To reduce source bias introduced by the training corpus itself, the system integrates multi-source open-source speech data. Overall, the training corpus consists of two parts. The first part is the speech deepfake corpus, which contains public speech deepfake detection datasets and covers major spoofing mechanisms, including text-to-speech (TTS), voice conversion (VC), neural vocoder reconstruction (NV), and neural codec resynthesis (NC). The second part is the supplementary bonafide corpus, which consists of additional real speech datasets and is used to expand the coverage of bonafide speech in terms of source, speaker, language, and recording condition. Table~\ref{tab:training_corpus} summarizes the main sources, languages, sample sizes, and generator metadata of the training corpus.

\definecolor{headergray}{gray}{0.92}
\definecolor{groupgray}{gray}{0.95}

\newcolumntype{Y}{>{\raggedright\arraybackslash}X}
\newcolumntype{R}[1]{>{\raggedleft\arraybackslash}p{#1}}
\newcolumntype{C}[1]{>{\centering\arraybackslash}p{#1}}

\begin{table*}[t]
    \centering
    \caption{Multi-source training corpus used in the proposed system. \textbf{\# Bonafide} and \textbf{\# Spoofed} denote the numbers of real and spoofed training utterances. \textbf{Generator Types} denotes coarse spoofing mechanisms, text-to-speech (TTS), voice conversion (VC), neural vocoder reconstruction (NV), and neural codec resynthesis (NC). \textbf{\# Generator} is the number of distinct spoof generators in metadata when annotated.}
    \label{tab:training_corpus}
    \small
    \setlength{\tabcolsep}{4.2pt}
    \renewcommand{\arraystretch}{1.12}

    \begin{tabularx}{\textwidth}{@{}Y C{1cm} C{1.5cm} R{1.45cm} R{1.45cm} C{2.25cm} C{1.62cm}@{}}
    \toprule
    \rowcolor{headergray}
    \textbf{Dataset} &
    \textbf{Year} &
    \textbf{Language} &
    \textbf{\# Bonafide} &
    \textbf{\# Spoofed} &
    \textbf{Generator Types} &
    \textbf{\# Generator} \\
    \midrule

    \rowcolor{groupgray}
    \multicolumn{7}{@{}c}{\textbf{\textit{Speech Deepfake Corpus}}} \\
    \addlinespace[1.5pt]

    ASVspoof2015~\citep{ASVspoof2015} & 2015 & EN & 9,404 & 184,000 & TTS, VC & 10 \\
    ASVspoof2019LA~\citep{wang2020asvspoof2019} & 2019 & EN & 2,580 & 22,800 & TTS, VC & 6 \\
    ASVspoof5~\citep{wang2024asvspoof5} & 2024 & EN & 18,797 & 163,560 & TTS, VC & 8 \\
    ADD2022~\citep{yi2022add} & 2022 & ZH & 3,312 & 24,772 & TTS, VC & -- \\
    ADD2023 Track1~\citep{yi2023add} & 2023 & ZH & -- & 24,072 & TTS, VC & -- \\
    FakeOrReal~\citep{for2019} & 2019 & EN & 26,900 & 26,900 & TTS & -- \\
    SpoofCeleb~\citep{jung2025spoofceleb} & 2024 & EN & 230,948 & 2,309,473 & TTS & 10 \\
    ReplayDF~\citep{muller2025replay} & 2025 & MULTI & 26,160 & 26,160 & TTS & 4 \\
    DFADD~\citep{du2024dfadd} & 2024 & EN & 44,455 & 163,500 & TTS & 5 \\
    MLAAD~\citep{muller2024mlaad} & 2024 & MULTI & -- & 172,393 & TTS & 58 \\
    LibriSeVoc~\citep{sun2023ai} & 2023 & EN & 13,201 & 79,206 & NV & 6 \\
    SpeechFake~\citep{huang2025speechfake} & 2025 & EN, ZH & 75,708 & 629,154 & TTS, VC, NV & 30 \\
    Wavefake~\citep{frank2021wavefake} & 2021 & EN, JA & 9,170 & 117,985 & NV & 6 \\
    CodecFake~\citep{xie2025codecfake} & 2024 & EN, ZH & 105,821 & 634,926 & NC & 7 \\

    \addlinespace[3pt]
    \rowcolor{groupgray}
    \multicolumn{7}{@{}c}{\textbf{\textit{Supplementary Bonafide Corpus}}} \\
    \addlinespace[1.5pt]

    LibriSpeech~\citep{panayotov2015librispeech} & 2015 & EN & 281,241 & -- & -- & -- \\
    AISHELL3~\citep{shi2020aishell} & 2020 & ZH & 88,035 & -- & -- & -- \\
    GigaSpeech~\citep{chen2021gigaspeech} & 2021 & EN & 885,397 & -- & -- & -- \\
    CNCeleb~\citep{fan2020cn} & 2019 & ZH & 524,787 & -- & -- & -- \\
    CommonVoice~\citep{ardila2020common} & 2020 & MULTI & 167,571 & -- & -- & -- \\

    \bottomrule
    \end{tabularx}
\end{table*}

\begin{itemize}
\item \textbf{Speech Deepfake Corpus.} The speech deepfake corpus brings together speech spoofing data from different development stages. Early data mainly come from ASVspoof2015 and ASVspoof2019LA, which are used to cover typical TTS and VC attacks in the logical access scenario. The training corpus then incorporates datasets such as ADD2022, ADD2023 Track1, FakeOrReal, SpoofCeleb, ReplayDF, and ASVspoof5 to supplement samples from more complex challenge protocols and noisy conditions. Furthermore, the system introduces DFADD, MLAAD~, LibriSeVoc, SpeechFake, Wavefake, and CodecFake to provide spoofing-related samples from more recent or more specific generation pipelines. These data mainly involve diffusion / flow-matching TTS, multilingual TTS, neural vocoder generation or reconstruction, and neural codec reconstruction, thereby extending the coverage of the training corpus to new generation and reconstruction pipelines.

\item \textbf{Supplementary Bonafide Corpus.} In addition to the bonafide speech included in the speech deepfake corpus, this system further introduces a supplementary bonafide corpus to enhance the distributional coverage of real speech. Recent studies~\cite{kwok2025bona} show that models trained on existing deepfake detection data can suffer performance degradation under complex real-speech conditions, and some bonafide speech may be misclassified as spoofed speech due to differences in noise, recording environment, or speaking style. Based on this observation, we explicitly increase the source diversity of bonafide speech during training to reduce the dependence of the model on limited real-speech conditions.

Specifically, the supplementary corpus is divided into two categories. The first category contains relatively standard read-speech recordings, including LibriSpeech and AISHELL3, which are used to supplement clean and large-scale bonafide speech. The second category contains real speech data from more open sources and more complex scenarios, including GigaSpeech, CommonVoice, and CNCeleb, which are used to introduce stronger variations in speaker, language, content, and recording condition.
\end{itemize}

\subsection{Attack- and Source-Balanced Data Sampling}

As shown in Table~\ref{tab:training_corpus}, different training data sources contribute substantially different numbers of utterances, and the number of attack generators covered on the spoofed speech side also varies. If all samples are directly merged and randomly sampled, the training process can be dominated by a small number of large-scale data sources or frequent attack generators. This may cause the model to overfit specific source distributions and produce clear performance trade-offs across different evaluation sets. Similar data mixing issues have also been widely discussed in multi-domain pretraining and multilingual learning~\cite{xie2023doremi,chungunimax,ye2025data}, where the sampling strategy directly affects the generalization ability and robustness of the model.

To reduce such data distribution bias, this system adopts attack- and source-balanced data sampling to construct the training distribution of each epoch. For spoofed speech, the system prioritizes the attack generator labels provided by the datasets as sampling units. If such labels are not available, the dataset itself is used as the sampling unit. Let $M$ denote the number of samples for each spoofing unit. In this system, we set $M=1000$. If there are $K$ spoofing units in total, the number of spoofed samples in the epoch is $MK$. The system then samples the same number of bonafide speech samples and distributes them uniformly across different real speech datasets. This strategy jointly balances bonafide and spoofed speech, attack generators on the spoofed side, and data sources on the bonafide side.

\subsection{Diverse Audio Augmentation}

In real applications, speech signals often undergo various processing procedures caused by recording devices, room acoustics, background noise, bandwidth limitations, platform compression, and network transmission. These factors may change the observable form of spoofing artifacts and may also make bonafide speech exhibit acoustic conditions different from those in the training set. To improve the robustness of the detector to changes in propagation chains, this system introduces diverse audio augmentation at the waveform level, thereby expanding the coverage of signal conditions during training.

\newcommand{\opcite}[2]{%
  \begin{tabular}[c]{@{}l@{}}
  #1 \\
  \citep{#2}
  \end{tabular}%
}

\newcolumntype{M}[1]{>{\raggedright\arraybackslash}m{#1}}

\begin{table*}[t]
    \centering
    \setlength{\belowcaptionskip}{6pt}
    \caption{Audio augmentation operators used during training. For each sample, the system applies augmentation with probability 0.5 by randomly selecting one or two operators.}
    \label{tab:audio_augmentation}
    \small
    \setlength{\tabcolsep}{5.0pt}
    \renewcommand{\arraystretch}{1.12}

    \begin{tabular}{@{}M{2.8cm} M{6.2cm} M{\dimexpr\textwidth-2.8cm-6.2cm-4\tabcolsep\relax}@{}}
    \toprule
    \rowcolor{headergray}
    \textbf{Operator} &
    \textbf{Setting} &
    \textbf{Simulated condition} \\
    \midrule

    \rowcolor{groupgray}
    \multicolumn{3}{@{}c}{\textbf{\textit{Acoustic- and Recording-related Augmentation}}} \\
    \addlinespace[1.5pt]
    
    \opcite{RawBoost}{tak2022rawboost}
    & Algorithms 1, 2, and 3
    & Device/channel nonlinearity, impulsive disturbance, and colored background noise. \\
    \addlinespace[0.3em]
    
    \opcite{RIR}{ko2017study}
    & Room impulse responses
    & Reverberant recordings in enclosed spaces. \\
    \addlinespace[0.3em]
    
    \opcite{MUSAN}{snyder2015musan}
    & \begin{tabular}[c]{@{}l@{}}
      Noise: 5--30 dB; Speech: 15--30 dB \\
      Music: 10--30 dB
      \end{tabular}
    & Additive background interference from environmental noise, competing speech, or music. \\
    \addlinespace[0.3em]

    Pitch Shift
    & Pitch offset: $[-1, +1]$ semitones
    & Small global pitch offsets across processing chains. \\
    \addlinespace[0.35em]

    \rowcolor{groupgray}
    \multicolumn{3}{@{}c}{\textbf{\textit{Transmission- and Platform-related Augmentation}}} \\
    \addlinespace[2pt]
    
    Filtering
    & \begin{tabular}[c]{@{}l@{}}
      Low-pass: 2.2--7.2 kHz; High-pass: 70--360 Hz \\
      Band-pass: 300--4000 Hz
      \end{tabular}
    & Narrowband or frequency-shaped transmission paths. \\
    \addlinespace[0.3em]
    
    Time Masking
    & One waveform segment with 1--10\% duration
    & Brief local dropouts or corrupted waveform segments. \\
    \addlinespace[0.3em]

    Codec Compression
    & \begin{tabular}[c]{@{}l@{}}
      MP3, Vorbis, Opus, AAC, Speex, GSM, G.711 \\
      G.723.1, and Encodec~\citep{defossez2022high}
      \end{tabular}
    & Lossy re-encoding by streaming, messaging, or upload pipelines. \\
    \addlinespace[0.3em]

    Packet Loss
    & Loss rate: 0.01--0.1; Packet length: 20 ms
    & VoIP packet loss with silence or repetition concealment. \\
    
    \bottomrule
    \end{tabular}
\end{table*}

\begin{itemize}
\item \textbf{Acoustic- and Recording-related Augmentation.} The system first considers variations introduced by acoustic environments and recording devices. RawBoost is used to simulate nonlinear channel responses, impulsive perturbations, and colored background noise. Pitch shift introduces slight global pitch changes, RIR convolution introduces room reverberation, and MUSAN mixes noise, speech, or music into the input speech with a random signal-to-noise ratio. These augmentations jointly cover source signals, device responses, spatial acoustics, and external interference, preventing the model from relying only on spoofing cues observed under clean close-talking conditions.

\item \textbf{Transmission- and Platform-related Augmentation.} The system further considers changes that may occur during speech transmission, uploading, forwarding, and platform processing. Filtering covers low-pass, high-pass, and band-pass conditions, corresponding to transmission paths with bandwidth limitations or frequency-response changes. Time masking simulates local waveform missingness or corruption. Codec compression covers multiple lossy speech or audio coding conditions, while packet loss simulates short packet drops and concealment strategies in VoIP transmission. This class of augmentations is used to improve the robustness of the model to propagation-chain changes and platform-processing distortions.
\end{itemize}

\section{Benchmark Evaluation}
\subsection{Evaluation Protocol}
This report uses Speech-DF-Arena~\cite{dowerah2026speech} as the main evaluation benchmark. The platform aggregates multiple public evaluation sets under a unified evaluation protocol, and we follow its reported results on 14 test sets to evaluate the general detection ability of the model. The evaluation sets cover the ASVspoof series, including ASVspoof19~\cite{wang2020asvspoof2019}, ASVspoof21-LA, ASVspoof21-DF~\cite{yamagishi2021asvspoof}, and ASVspoof24-Eval~\cite{wang2024asvspoof5}; the ADD challenge series, including ADD 2022 Track 1 (ADD22-T1), ADD 2022 Track 3 (ADD22-T3)~\cite{yi2022add}, ADD 2023 Round 1 (ADD23-R1), and ADD 2023 Round 2 (ADD23-R2)~\cite{yi2023add}; as well as In-the-Wild (ITW)~\cite{muller2022does}, FakeOrReal (FoR)~\cite{for2019}, CodecFake (CF)~\cite{xie2025codecfake}, LibriSeVoc (LSV)~\cite{sun2023ai}, DFADD~\cite{du2024dfadd}, and SONAR~\cite{li2410sonar}. These test sets span diverse corpus compositions, spoofing generation pipelines, and signal processing conditions.

It should be noted that the training corpus used in this report includes the official training partitions of some benchmarks in the evaluation suite. However, many of these benchmarks have clear train--test distribution gaps in spoofing methods, generation models, and other signal-processing conditions. Therefore, the evaluation does not reduce to a purely in-distribution test and still provides evidence of the model's ability to generalize to unseen attacks.

Speech-DF-Arena uses EER, ACC, and F1-score as the main evaluation metrics, and reports dataset-level, pooled, and average results separately. Dataset-level results are computed independently on each test set. Pooled results merge all scores, determine a unified threshold, and compute the metrics accordingly; these results are used as the basis for leaderboard ranking. Average results are obtained by directly averaging the dataset-level results across all test sets.

\subsection{Leaderboard Performance}

Table~\ref{tab:leaderboard_eer} compares the EER performance of Teffic-Audio with the public systems in the current public snapshot of the official Speech-DF-Arena leaderboard. Complete ACC and F1-score results are provided in Appendix Tables~\ref{tab:leaderboard_acc} and~\ref{tab:leaderboard_f1}. In terms of pooled EER, Teffic-Audio achieves $1.454\%$, which is lower than all currently public systems. Compared with the top three systems, Modulate-VELMA-2-Synthetic-Voice~\cite{modulate_velma2_synthetic_voice_2026} ($1.586\%$), Resemble-Detect-3B-Omni~\cite{resemble_detect_3b_omni_2026} ($2.099\%$), and Hiya-Authenticity-Verification-Multi-v1~\cite{hiya_authenticity_verification_multi_v1_2026} ($2.324\%$), Teffic-Audio reduces the pooled EER by approximately $8.3\%$, $30.7\%$, and $37.4\%$ in relative terms, respectively. This result indicates that, under a mainstream detector architecture, the proposed training recipe can provide the detection system with strong cross-domain discriminative ability.

From the perspective of individual test sets, Teffic-Audio achieves the lowest EER on five test sets, including CodecFake, ADD 2022 Track 3, ADD 2023 R1, DFADD~, and LibriSeVoc, and maintains low error rates on challenging test sets such as ADD 2022 Track 1 and ASV2024-Eval. This shows that the performance of the system is not supported only by advantages on a few test sets, but instead remains stable across multiple types of evaluation conditions. When model size is further considered, Teffic-Audio has 590.0M parameters, which is only larger than Modulate-VELMA-2-Synthetic-Voice and is substantially smaller than Resemble-Detect-3B-Omni (3B) and Hiya-Authenticity-Verification-Multi-v1 (1B). This suggests that the proposed system not only achieves strong spoofing detection performance, but also presents a favorable performance--complexity trade-off.

\definecolor{oursrow}{HTML}{E6F4EF}

\newlength{\syscellheight}

\newcommand{\sysone}[1]{%
  \parbox[c][\syscellheight][c]{\linewidth}{\raggedright #1}%
}

\newcommand{\systwo}[2]{%
  \parbox[c][\syscellheight][c]{\linewidth}{%
    \raggedright #1\\[-0.2ex]#2%
  }%
}

\newcommand{\syscite}[2]{%
  \systwo{#1}{\citep{#2}}%
}

\begin{table*}[!t]
    \centering
    \caption{EER(\%) performance on Speech-DF-Arena. The table reports Teffic-Audio and all systems from the current public leaderboard snapshot, covering licence, parameter size, overall and dataset-level EER results on the 14 evaluation sets. Params denotes the number of parameters in millions. Pooled and Avg. denote pooled EER and the average of dataset-level EERs, respectively. Systems are ranked by pooled EER.}
    \label{tab:leaderboard_eer}
    \tiny
    \setlength{\syscellheight}{2\baselineskip}
    \setlength{\tabcolsep}{2.2pt}
    \resizebox{\linewidth}{!}{%
    \begin{tabular}{p{2.75cm}|*{4}{c}|*{14}{c}}
    \toprule
    \rowcolor{headergray}
     & & & & & &
    \multicolumn{4}{c}{\textbf{ASVspoof}} &
     & &
    \multicolumn{4}{c@{}}{\textbf{ADD}} &
     & & \\
    \noalign{\global\aboverulesep=0pt\global\belowrulesep=0pt}
    \cmidrule(lr){7-10}\cmidrule(lr){13-16}
    \noalign{\global\aboverulesep=.4ex\global\belowrulesep=.65ex}
    \rowcolor{headergray}
    \multirow{-2}{*}{\textbf{System}} &
    \multirow{-2}{*}{\textbf{Licence}} &
    \multirow{-2}{*}{\textbf{Params}} &
    \multirow{-2}{*}{\textbf{Pooled}} &
    \multirow{-2}{*}{\textbf{Avg.}} &
    \multirow{-2}{*}{\textbf{ITW}} &
    \multicolumn{1}{c}{\textbf{19}} &
    \multicolumn{1}{c}{\textbf{21-LA}} &
    \multicolumn{1}{c}{\textbf{21-DF}} &
    \multicolumn{1}{c}{\textbf{24-E}} &
    \multicolumn{1}{c}{\multirow{-2}{*}{\textbf{FoR}}} &
    \multicolumn{1}{c}{\multirow{-2}{*}{\textbf{CF}}} &
    \multicolumn{1}{c}{\textbf{22-T1}} &
    \multicolumn{1}{c}{\textbf{22-T3}} &
    \multicolumn{1}{c}{\textbf{23-R1}} &
    \multicolumn{1}{c}{\textbf{23-R2}} &
    \multirow{-2}{*}{\textbf{DFADD}} &
    \multirow{-2}{*}{\textbf{LSV}} &
    \multirow{-2}{*}{\textbf{SONAR}} \\
    \midrule

    \rowcolor{oursrow}
    \sysone{Teffic-Audio} & Proprietary & 590.0 & \textbf{1.454} & 1.236 & 1.321 & 0.242 & 2.038 & 0.262 & 1.405 & 0.972 & \textbf{0.748} & 7.031 & \textbf{0.837} & \textbf{0.356} & 1.836 & \textbf{0.000} & \textbf{0.000} & 0.251 \\

    \midrule

    \systwo{Modulate-VELMA-2-Synthetic-}{Voice~\citep{modulate_velma2_synthetic_voice_2026}}
    & Proprietary & 316.0 & 1.586 & \textbf{1.104} & 1.271 & 0.299 & 1.330 & 0.331 & \textbf{0.384} & 0.133 & 1.538 & \textbf{5.059} & 1.174 & 1.041 & \textbf{1.742} & 0.000 & 0.265 & 0.888 \\

    \syscite{Resemble-Detect-3B-Omni}{resemble_detect_3b_omni_2026}
    & Proprietary & 3000.0 & 2.099 & 2.570 & 1.347 & 2.531 & 3.017 & 0.579 & 0.453 & 2.341 & 2.689 & 9.958 & 1.549 & 3.390 & 7.360 & 0.100 & \textbf{0.000} & 1.139 \\

    \systwo{Hiya-Authenticity-Verification}{Multi-v1~\citep{hiya_authenticity_verification_multi_v1_2026}}
    & Proprietary & 1000.0 & 2.324 & 2.113 & 0.667 & 0.301 & 1.006 & 1.318 & 0.787 & \textbf{0.000} & 5.733 & 12.099 & 1.188 & 1.976 & 4.006 & 0.017 & 0.003 & 0.481 \\

    \syscite{DLMSL-SpeakSure-v0.1}{dlmsl_speaksure_v01_2026}
    & Proprietary & 658.6 & 6.142 & 3.954 & 1.278 & \textbf{0.042} & \textbf{0.081} & \textbf{0.013} & 12.894 & 0.265 & 6.828 & 15.140 & 2.373 & 5.654 & 8.280 & 0.133 & 0.120 & \textbf{0.074} \\

    \sysone{Whispeak~\citep{whispeak_2026}}
    & Proprietary & 98.9 & 8.060 & 3.049 & 1.268 & 0.394 & 3.578 & 3.235 & 9.924 & 1.016 & 0.856 & 11.942 & 2.310 & 2.618 & 5.007 & \textbf{0.000} & 0.044 & 0.533 \\

    \sysone{DF-Raptor~\citep{df_raptor_2026}}
    & Proprietary & 100.0 & 8.350 & 8.390 & 3.191 & 7.695 & 12.865 & 9.288 & 8.036 & 2.694 & 12.731 & 28.882 & 3.493 & 6.452 & 9.178 & 1.867 & 3.442 & 7.784 \\

    \syscite{DF\_Arena\_1B\_V\_1}{kulkarni2026compactsslbackbonesmatter}
    & Open & 1000.0 & 9.524 & 5.919 & 0.906 & 1.139 & 4.657 & 1.749 & 17.250 & 2.915 & 8.369 & 22.210 & 2.204 & 5.082 & 11.544 & \textbf{0.000} & 0.151 & 1.087 \\

    \sysone{Syntra Detector~\citep{syntra_detector_2026}}
    & Proprietary & 584.0 & 10.764 & 6.109 & 3.978 & 1.482 & 14.053 & 2.026 & 15.960 & 0.530 & 1.187 & 23.582 & 2.985 & 3.653 & 9.604 & \textbf{0.000} & 1.171 & 0.658 \\

    \syscite{DF\_Arena\_500M\_V\_1}{kulkarni2026compactsslbackbonesmatter}
    & Open & 500.0 & 10.880 & 5.780 & 1.760 & 1.090 & 4.230 & 3.300 & 12.390 & 2.300 & 6.360 & 23.980 & 2.770 & 7.470 & 12.300 & \textbf{0.000} & 0.120 & 1.900 \\

    \sysone{MoLEx~\citep{pan2025molex}}
    & Proprietary & 376.4 & 12.405 & 9.519 & \textbf{0.034} & 0.284 & 6.325 & 1.885 & 15.870 & 0.177 & 32.401 & 31.939 & 3.663 & 11.111 & 19.021 & 6.462 & 0.303 & 0.910 \\

    \syscite{Resemble Detect}{resemble_detect_2026}
    & Proprietary & 2112.0 & 12.747 & 10.830 & 3.946 & 1.321 & 1.647 & 3.793 & 16.291 & 1.369 & 33.044 & 28.218 & 6.113 & 21.073 & 28.278 & \textbf{0.000} & 1.628 & 2.989 \\

    \sysone{DF\_Arena\_100M\_V\_1}
    & Closed & 100.0 & 13.921 & 8.398 & 2.218 & 1.535 & 7.612 & 5.501 & 21.398 & 7.420 & 8.749 & 27.063 & 5.421 & 8.904 & 17.041 & \textbf{0.000} & 0.221 & 4.483 \\

    \sysone{DF\_Arena\_100M\_V\_0}
    & Closed & 100.0 & 15.930 & 10.120 & 4.290 & 4.160 & 10.060 & 8.830 & 21.280 & 6.140 & 10.390 & 31.960 & 6.690 & 12.690 & 19.010 & 0.130 & 0.420 & 5.730 \\

    \sysone{XLSR+SLS~\citep{zhang2024audio}}
    & Open & 340.0 & 16.079 & 14.015 & 7.455 & 0.231 & 2.869 & 1.910 & 18.764 & 5.080 & 33.437 & 33.950 & 15.744 & 19.374 & 21.096 & 7.542 & 1.969 & 24.723 \\

    \sysone{TCM~\citep{truong2024temporal}}
    & Open & 319.0 & 16.691 & 15.846 & 7.794 & 0.188 & 2.996 & 2.145 & 18.850 & 10.689 & 36.008 & 37.403 & 20.942 & 23.427 & 22.743 & 8.887 & 2.348 & 26.572 \\

    \syscite{BiCrossMamba-ST}{kheir2025bicrossmamba}
    & Open & 318.2 & 17.154 & 15.778 & 7.937 & 0.707 & 3.827 & 2.347 & 13.669 & 6.846 & 37.703 & 30.443 & 18.685 & 29.444 & 29.929 & 8.505 & 2.120 & 27.356 \\

    \sysone{Nes2NetX~\citep{liu2025nes2net}}
    & Open & 317.9 & 17.366 & 16.186 & 7.751 & 0.122 & 2.173 & 1.493 & 22.060 & 6.316 & 39.342 & 34.471 & 26.555 & 21.135 & 18.450 & 11.146 & 2.878 & 31.536 \\

    \syscite{Wav2Vec2 AASIST}{tak2022automatic}
    & Open & 317.8 & 19.607 & 18.138 & 11.196 & 0.221 & 0.823 & 6.630 & 16.247 & 7.465 & 43.368 & 31.050 & 16.528 & 27.742 & 21.929 & 11.927 & 11.208 & 46.124 \\

    \syscite{XLSR Mamba}{xiao2025xlsr}
    & Open & 319.0 & 20.591 & 14.647 & 6.709 & 0.421 & 0.931 & 1.884 & 14.401 & 6.714 & 35.266 & 34.228 & 19.368 & 21.848 & 20.151 & 10.698 & 2.234 & 24.264 \\

    \syscite{Whisper Mesonet}{kawa2023improved}
    & Open & 7.6 & 23.551 & 28.355 & 26.725 & 5.833 & 15.821 & 2.111 & 22.549 & 47.747 & 34.661 & 38.383 & 24.008 & 41.250 & 44.562 & 24.120 & 15.335 & 58.663 \\

    \syscite{Wav2Vec2 ECAPA}{kulkarni2024exploring}
    & Open & 324.0 & 28.995 & 37.922 & 30.696 & 29.695 & 26.606 & 22.438 & 18.656 & 62.323 & 40.975 & 46.435 & 21.871 & 35.286 & 36.702 & 75.066 & 28.515 & 64.566 \\

    \sysone{AASIST~\citep{jung2022aasist}}
    & Open & 0.3 & 32.771 & 34.404 & 43.010 & 0.830 & 11.461 & 21.071 & 35.534 & 21.643 & 51.058 & 47.919 & 33.187 & 47.757 & 32.469 & 41.860 & 38.019 & 57.471 \\

    \syscite{WavLM ECAPA}{kulkarni2024exploring}
    & Open & 102.0 & 33.463 & 28.994 & 34.649 & 0.761 & 6.675 & 15.945 & 25.992 & 23.366 & 46.185 & 44.169 & 39.411 & 29.000 & 31.937 & 29.535 & 31.882 & 41.944 \\

    \sysone{RawGatST~\citep{tak2021end_2}}
    & Open & 0.4 & 33.553 & 34.821 & 52.538 & 1.060 & 10.253 & 23.262 & 40.291 & 53.092 & 50.000 & 42.903 & 32.302 & 37.866 & 27.337 & 23.704 & 43.364 & 50.784 \\

    \sysone{RawTFNet~\citep{xiao2025rawtfnet}}
    & Open & 0.2 & 39.938 & 32.820 & 38.727 & 1.890 & 5.041 & 16.820 & 44.740 & 36.572 & 51.776 & 43.850 & 36.807 & 38.726 & 30.539 & 22.508 & 29.578 & 54.786 \\

    \syscite{Hubert ECAPA}{kulkarni2024exploring}
    & Open & 102.0 & 43.156 & 33.843 & 38.658 & 1.058 & 12.554 & 13.793 & 31.396 & 33.746 & 46.224 & 47.661 & 39.084 & 49.569 & 43.952 & 34.568 & 32.033 & 40.198 \\

    \sysone{Rawnet2~\citep{tak2021end}}
    & Open & 17.6 & 45.996 & 47.567 & 49.187 & 33.039 & 40.071 & 40.669 & 41.219 & 65.680 & 50.219 & 49.359 & 47.627 & 55.641 & 64.554 & 39.070 & 48.167 & 43.009 \\

    \bottomrule
    \end{tabular}
    }%
\end{table*}

\section{Ablation Analysis}

\subsection{Effect of the Training Recipe}
\label{subsec:ablation_training_recipe}
Table~\ref{tab:ablation_recipe_eer} presents the impact of different training strategies on system performance under the fixed architecture. The baseline simply merges multiple speech deepfake datasets and applies random sampling during training, yielding a pooled EER of 7.159\% and an average EER of 3.693\%. The clear gap between this baseline and the final system indicates that merely increasing the diversity of training data sources is insufficient for robust cross-dataset generalization. After introducing attack- and source-balanced sampling, the pooled EER decreases to 6.456\%. This suggests that explicitly balancing attack types and data sources can alleviate distributional bias in multi-source training. With the further inclusion of the supplementary bonafide corpus, the pooled EER is reduced to 5.435\%. This result demonstrates that a general-purpose detector requires diverse spoofed samples as well as broad coverage of bonafide speech distributions.

For data augmentation, we compare two schemes: using RawBoost alone and using the Diverse Audio Augmentation strategy adopted in this work. Compared with the system after incorporating the supplementary bonafide corpus, using RawBoost alone does not yield consistent or substantial improvements in the overall metrics. A closer examination of individual test sets shows that RawBoost brings clear gains mainly on scenarios such as ASVspoof 2021-LA and ASVspoof 2021-DF. However, it provides limited improvement or even causes degradation on datasets like ASVspoof 2024-Eval. This suggests that RawBoost can simulate certain channel and noise perturbations as a commonly used waveform-level augmentation method in speech deepfake detection. Nevertheless, it is insufficient to consistently improve the cross-dataset generalization of a general-purpose detector under a comprehensive evaluation setting.

In contrast, Diverse Audio Augmentation reduces the pooled EER to 1.454\% and the average EER to 1.236\%, making it a key factor in improving system generalization. The performance gains are particularly pronounced on test sets that are more sensitive to coding compression and channel variations. Specifically, the EER on ASVspoof 2024-Eval decreases from 17.814\% to 1.405\%, while that on ADD 2022-Track1 decreases from 13.794\% to 7.031\%. These results indicate that richer audio augmentation can substantially improve the model's adaptability to complex acoustic conditions. At the same time, no systematic degradation is observed on other test sets. This suggests that the proposed strategy does not weaken the model's ability to discriminate spoofing cues under conventional speech conditions, thereby leading to stronger cross-dataset robustness.

\begin{table*}[t]
\centering
\caption{Ablation on the training recipe. Architecture is fixed to the w2v-BERT 2.0 encoder with MHASP throughout. \textbf{Baseline} uses merged speech deepfake corpus with random sampling only.}
\label{tab:ablation_recipe_eer}
\tiny
\setlength{\tabcolsep}{2.2pt}
\resizebox{\linewidth}{!}{%
\begin{tabular}{p{3.5cm}|*{3}{c}|*{14}{c}}
\toprule
\rowcolor{headergray}
 & & & & &
\multicolumn{4}{c}{\textbf{ASVspoof}} &
 & &
\multicolumn{4}{c}{\textbf{ADD}} &
 & & \\
\noalign{\global\aboverulesep=0pt\global\belowrulesep=0pt}
\cmidrule(lr){6-9}\cmidrule(lr){12-15}
\noalign{\global\aboverulesep=.4ex\global\belowrulesep=.65ex}
\rowcolor{headergray}
\multirow{-2}{*}{\textbf{Configuration}} &
\multirow{-2}{*}{\textbf{Params}} &
\multirow{-2}{*}{\textbf{Pooled}} &
\multirow{-2}{*}{\textbf{Avg.}} &
\multirow{-2}{*}{\textbf{ITW}} &
\multicolumn{1}{c}{\textbf{19}} &
\multicolumn{1}{c}{\textbf{21-LA}} &
\multicolumn{1}{c}{\textbf{21-DF}} &
\multicolumn{1}{c}{\textbf{24-E}} &
\multicolumn{1}{c}{\multirow{-2}{*}{\textbf{FoR}}} &
\multicolumn{1}{c}{\multirow{-2}{*}{\textbf{CF}}} &
\multicolumn{1}{c}{\textbf{22-T1}} &
\multicolumn{1}{c}{\textbf{22-T3}} &
\multicolumn{1}{c}{\textbf{23-R1}} &
\multicolumn{1}{c}{\textbf{23-R2}} &
\multirow{-2}{*}{\textbf{DFADD}} &
\multirow{-2}{*}{\textbf{LSV}} &
\multirow{-2}{*}{\textbf{SONAR}} \\
\midrule

Baseline & 590.0 & 7.159 & 3.693 & 1.445 & 0.600 & 5.825 & 1.619 & 19.650 & 0.345 & \textbf{0.097} & 14.885 & 1.915 & 1.185 & 3.655 & 0.017 & \textbf{0.000} & 0.459 \\
\midrule
\hspace*{0.6em}+ Attack- \& Source-balanced Sampling & 590.0 & 6.456 & 3.539 & \textbf{1.173} & 0.243 & 6.822 & 0.540 & 17.840 & 0.216 & 0.144 & 14.730 & 1.352 & 1.502 & 4.603 & \textbf{0.000} & \textbf{0.000} & 0.377 \\
\hspace*{1.2em}+ Supplementary Bonafide Corpus & 590.0 & 5.435 & 3.289 & 1.242 & 0.598 & 7.972 & 1.304 & 13.630 & \textbf{0.193} & 0.235 & 14.038 & 1.557 & 0.581 & 3.509 & \textbf{0.000} & \textbf{0.000} & 1.183 \\
\hspace*{1.8em}+ RawBoost Augmentation & 590.0 & 5.896 & 3.081 & 1.187 & \textbf{0.178} & 4.552 & 0.347 & 17.814 & 0.410 & 0.268 & 13.794 & 1.093 & 0.439 & 2.046 & \textbf{0.000} & \textbf{0.000} & 1.006 \\
\hspace*{1.8em}+ Diverse Audio Augmentation & 590.0 & \textbf{1.454} & \textbf{1.236} & 1.321 & 0.242 & \textbf{2.038} & \textbf{0.262} & \textbf{1.405} & 0.972 & 0.748 & \textbf{7.031} & \textbf{0.837} & \textbf{0.356} & \textbf{1.836} & \textbf{0.000} & \textbf{0.000} & \textbf{0.251} \\

\bottomrule
\end{tabular}
}%
\end{table*}

\subsection{Effect of the Model Architecture}
\label{subsec:ablation_architecture}

Table~\ref{tab:ablation_architecture} presents the ablation study on model architecture under our training recipe. We examine three aspects of the detector design: the choice of speech encoder, the utterance-level pooling layer, and the encoder depth of Teffic-Audio.


\begin{itemize}
    \item \textbf{Encoder Backbone.} 
    We first compare the impact of different encoder backbones on system performance. Although traditional convolutional models such as AASIST~\cite{jung2022aasist} and Res2Net~\cite{li2021replay} have relatively small parameter sizes, they perform substantially worse than detectors based on large-scale SSL-initialized encoders in the comprehensive evaluation setting. Introducing such encoders leads to clear performance improvements. For example, WavLM Large~\cite{chen2022wavlm} + MHASP and XLS-R 1B~\cite{babu2021xls} + MHASP achieve pooled EERs of 4.332\% and 4.419\%, respectively, outperforming AASIST and Res2Net. 

    A further comparison between different scales of the same encoder family shows that WavLM Base + MHASP and XLS-R 300M + MHASP obtain pooled EERs of 9.446\% and 6.079\%, respectively, both significantly higher than their corresponding large-scale variants. This indicates that increasing the encoder scale generally helps improve system performance. Among different SSL-initialized backbones, the w2v-BERT 2.0 encoder adopted in Teffic-Audio achieves the best result, further reducing the pooled EER to 1.454\%. It also shows stronger generalization across multiple test sets. These results suggest that the performance of general-purpose speech deepfake detection is not determined solely by parameter size. It is also closely related to the encoder's pretraining objective and acoustic representation capability.

    From another perspective, under the same evaluation protocol, models trained with our recipe also substantially outperform leaderboard systems that use similar backbones. Specifically, AASIST achieves a pooled EER of 15.668\%, corresponding to a 52.2\% reduction compared with the leaderboard AASIST system. WavLM Base + MHASP achieves 9.446\%, reducing the EER by 71.7\% compared with WavLM-ECAPA~\citep{kulkarni2024exploring}. XLS-R 300M + MHASP achieves 6.079\%, reducing the EER by 62.2\% compared with XLSR+SLS~\citep{zhang2024audio}. These results further indicate that the proposed training recipe can effectively improve the overall performance of different encoder backbones in general-purpose deepfake detection.
    
    \item \textbf{Pooling Layer.} 
    We next compare the impact of different utterance-level pooling layers. With the w2v-BERT 2.0 encoder fixed, mean pooling and ASP achieve pooled EERs of 2.269\% and 2.339\%, respectively. Both yield higher EERs than the final MHASP-based system, indicating that the choice of pooling layer also affects overall system performance. Notably, although ASP introduces attention-based pooling, it does not improve over mean pooling. This suggests that simple frame-level weighting may be insufficient to exploit the fine-grained representations produced by the encoder. In contrast, MHASP aggregates frame-level statistics from different subspaces through multiple attention heads, which provides a more effective utterance-level representation for cross-dataset generalization.

    \item \textbf{Encoder Depth.} 
    The preceding backbone comparison shows that directly adopting smaller SSL backbones leads to clear performance degradation. We therefore further examine whether the Teffic-Audio encoder can be made shallower to reduce parameter size while maintaining stable cross-dataset generalization. To this end, we train and evaluate variants with different numbers of Conformer blocks. The results show that overall performance improves as encoder depth increases. From $N=3$ to the full 24-layer configuration, the pooled EER gradually decreases from 4.735\% to 1.454\%. This result indicates that, after selecting an effective backbone, encoder depth remains a key factor in the trade-off between performance and complexity.

    Further examination on individual test sets shows that the benefits of increasing depth are not uniform across all evaluation scenarios. On test sets such as ITW, ASVspoof21-DF, DFADD, and LSV, the 4-layer or 6-layer configurations already achieve relatively low EERs. This suggests that spoofing cues in many milder test conditions can already be captured by a shallower encoder. In contrast, the advantage of the full-depth model is mainly observed on more challenging test sets, such as ASVspoof24-E and ADD22-T1. For example, the EER on ASVspoof24-E decreases from 6.744\% with $N=4$ to 1.405\% with the full model. The EER on ADD22-T1 decreases from 10.734\% to 7.031\%. These results suggest that increasing encoder depth mainly improves the model's adaptability to challenging scenarios. Meanwhile, a clear degradation is observed when reducing the depth from $N=4$ to $N=3$. The pooled EER increases from 3.346\% to 4.735\%, corresponding to a relative increase of 41.5\%. This indicates that overly reducing the encoder depth weakens the stability of cross-dataset discrimination.

    It is worth noting that shallower Teffic-Audio configurations still retain strong performance. The 4-layer configuration contains only 106.4M parameters, yet achieves a pooled EER of 3.346\%. This outperforms WavLM Base~\cite{chen2022wavlm} + MHASP, which has a comparable parameter size. The 6-layer configuration further achieves a pooled EER of 2.839\% with 154.8M parameters. This performance is close to leading systems on the leaderboard, such as Resemble-Detect-3B-Omni and Hiya-Authenticity-Verification-Multi-v1. These results indicate that reducing the depth of an effective backbone provides a better performance--complexity trade-off than directly replacing it with a smaller SSL encoder. Combined with an appropriate training recipe, shallower configurations can still maintain strong cross-dataset generalization.

\end{itemize}

\begin{table*}[t]
    \centering
    \caption{Ablation study on model architecture. Teffic-Audio refers to the full system using MHASP and a 24-layer encoder initialized from w2v-BERT 2.0. All metric columns report EER (\%). MP and ASP denote mean pooling and attentive statistics pooling, respectively. For Teffic-Audio depth variants, $N$ denotes the encoder depth, and the $N$-layer encoders are initialized from the first $N$ Conformer blocks of w2v-BERT 2.0.}
    \label{tab:ablation_architecture}
    \tiny
    \setlength{\tabcolsep}{2.2pt}
    \resizebox{\linewidth}{!}{%
    \begin{tabular}{p{2.5cm}|*{3}{c}|*{14}{c}}
    \toprule
    \rowcolor{headergray}
     & & & & &
    \multicolumn{4}{c}{\textbf{ASVspoof}} &
     & &
    \multicolumn{4}{c}{\textbf{ADD}} &
     & & \\
    \noalign{\global\aboverulesep=0pt\global\belowrulesep=0pt}
    \cmidrule(lr){6-9}\cmidrule(lr){12-15}
    \noalign{\global\aboverulesep=.4ex\global\belowrulesep=.65ex}
    \rowcolor{headergray}
    \multirow{-2}{*}{\textbf{Configuration}} &
    \multirow{-2}{*}{\textbf{Params}} &
    \multirow{-2}{*}{\textbf{Pooled}} &
    \multirow{-2}{*}{\textbf{Avg.}} &
    \multirow{-2}{*}{\textbf{ITW}} &
    \multicolumn{1}{c}{\textbf{19}} &
    \multicolumn{1}{c}{\textbf{21-LA}} &
    \multicolumn{1}{c}{\textbf{21-DF}} &
    \multicolumn{1}{c}{\textbf{24-E}} &
    \multicolumn{1}{c}{\multirow{-2}{*}{\textbf{FoR}}} &
    \multicolumn{1}{c}{\multirow{-2}{*}{\textbf{CF}}} &
    \multicolumn{1}{c}{\textbf{22-T1}} &
    \multicolumn{1}{c}{\textbf{22-T3}} &
    \multicolumn{1}{c}{\textbf{23-R1}} &
    \multicolumn{1}{c}{\textbf{23-R2}} &
    \multirow{-2}{*}{\textbf{DFADD}} &
    \multirow{-2}{*}{\textbf{LSV}} &
    \multirow{-2}{*}{\textbf{SONAR}} \\
    \midrule

    Teffic-Audio & 590.0 & \textbf{1.454} & \textbf{1.236} & 1.321 & \textbf{0.242} & \textbf{2.038} & 0.262 & \textbf{1.405} & 0.972 & 0.748 & 7.031 & 0.837 & \textbf{0.356} & \textbf{1.836} & \textbf{0.000} & \textbf{0.000} & \textbf{0.251} \\
    \midrule

    \rowcolor{groupgray}
    \multicolumn{1}{p{2.5cm}}{} & \multicolumn{17}{c}{\textbf{\textit{Encoder Backbone}}} \\
    \addlinespace[1.5pt]
    
    AASIST & 0.3 & 15.668 & 11.480 & 2.225 & 16.228 & 20.644 & 11.346 & 15.896 & 9.189 & 13.504 & 29.852 & 6.730 & 9.827 & 16.002 & 0.664 & 1.325 & 7.294 \\
    Res2Net + MHASP & 2.8 & 15.023 & 10.744 & 8.847 & 4.639 & 21.298 & 23.619 & 22.809 & 11.734 & 0.864 & 27.747 & 4.952 & 5.109 & 8.341 & \textbf{0.000} & 0.189 & 10.261  \\
    WavLM Base + MHASP & 117.7 & 9.446 & 8.122 & 3.449 & 6.254 & 12.162 & 4.957 & 4.530 & 6.042 & 10.854 & 30.250 & 5.275 & 8.514 & 18.639 & 0.133 & 0.032 & 2.611 \\
    WavLM Large + MHASP & 315.4 & 4.332 & 4.005 & 1.700 & 2.083 & 7.570 & 2.246 & 4.764 & 4.489 & 1.903 & 14.158 & 2.143 & 5.503 & 8.274 & \textbf{0.000} & \textbf{0.000} & 1.235 \\
    XLS-R 300M + MHASP & 339.8 & 6.079 & 4.519 & 2.074 & 2.294 & 8.672 & 5.104 & 5.101 & 4.142 & 2.836 & 15.581 & 1.985 & 5.010 & 9.102 & \textbf{0.000} & 0.006 & 1.361 \\
    XLS-R 1B + MHASP & 962.5 & 4.419 & 2.909 & 1.375 & 0.493 & 2.249 & 1.129 & 7.877 & 1.552 & 2.164 & 14.877 & 2.327 & 1.413 & 4.909 & \textbf{0.000} & \textbf{0.000} & 0.355 \\

    \rowcolor{groupgray}
    \multicolumn{1}{p{2.5cm}}{} & \multicolumn{17}{c}{\textbf{\textit{Pooling Layer}}} \\
    \addlinespace[1.5pt]

    Teffic-Audio w/ MP & 587.9 & 2.269 & 1.515 & 1.234 & 0.244 & 2.666 & \textbf{0.255} & 4.727 & \textbf{0.474} & 0.898 & 6.821 & 0.769 & 0.436 & 2.057 & \textbf{0.000} & \textbf{0.000} & 0.459 \\
    Teffic-Audio w/ ASP & 590.5 & 2.339 & 1.499 & 1.036 & 0.247 & 3.036 & 0.264 & 4.514 & 0.798  & 0.710 & \textbf{6.803} & \textbf{0.712} & 0.443 & 1.946 & \textbf{0.000} & \textbf{0.000} & 0.481 \\

    \rowcolor{groupgray}
    \multicolumn{1}{p{2.5cm}}{} & \multicolumn{17}{c}{\textbf{\textit{Encoder Depth}}} \\
    \addlinespace[1.5pt]
    
    Teffic-Audio ($N{=}3$) & 82.2 & 4.735 & 3.280 & 1.459 & 2.423 & 6.449 & 1.835 & 9.901 & 2.785 & \textbf{0.594} & 13.075 & 1.340 & 0.798 & 3.683 & \textbf{0.000} & \textbf{0.000} & 1.582 \\
    Teffic-Audio ($N{=}4$) & 106.4 & 3.346 & 2.415 & 1.083 & 0.699 & 4.244 & 0.593 & 6.744 & 4.726 & 0.595 & 10.734 & 1.059 & 0.440 & 2.230 & \textbf{0.000} & \textbf{0.000} & 0.658 \\
    Teffic-Audio ($N{=}6$) & 154.8 & 2.839 & 2.005 & \textbf{0.922} & 0.515 & 3.134 & 0.484 & 5.306 & 2.633 & 0.937 & 9.640 & 0.918 & 0.526 & 2.153 & \textbf{0.000} & \textbf{0.000} & 0.910 \\
    Teffic-Audio ($N{=}12$) & 299.9 & 2.317 & 1.965 & 1.030 & 0.565 & 3.307 & 0.455 & 3.769 & 4.057 & 1.279 & 8.754 & 0.826 & 0.576 & 2.324 & \textbf{0.000} & \textbf{0.000} & 0.564 \\
    \bottomrule
    \end{tabular}
    }%
\end{table*}


\section{Conclusion}

This report presented Teffic-Audio, a general speech deepfake detection system for comprehensive evaluation environment. The system follows a standard detection pipeline and focuses on the construction of an effective training distribution. Trained only with open-source data, Teffic-Audio achieves a pooled EER of 1.454\% on the 14 test sets of Speech-DF-Arena and ranks first compared with all currently public systems.
Ablation studies show that the final performance cannot be attributed merely to large-scale data aggregation. Balanced sampling, supplementary bonafide speech, and especially diverse audio augmentation all contribute to stronger cross-dataset generalization. The architecture analysis further shows that the encoder backbone and pooling layer have clear effects on system performance. In particular, reducing the depth of the system backbone still yields competitive results while substantially decreasing the number of parameters. These findings show that Teffic-Audio provides an effective and practical baseline for general speech deepfake detection under comprehensive evaluation environment.

\section*{Authors}

\textbf{Contributors:} Wan Lin, Li Wang, Jindong Wang, Kunyu Feng, Zhizheng Wu.

\bibliographystyle{unsrtnat}
\bibliography{references}

@article{dowerah2026speech,
  title={Speech df arena: A leaderboard for speech deepfake detection models},
  author={Dowerah, Sandipana and Kulkarni, Atharva and Kulkarni, Ajinkya and Tran, Hoan My and Kalda, Joonas and Fedorchenko, Artem and Fauve, Benoit and Lolive, Damien and Alum{\"a}e, Tanel and Doss, Mathew Magimai-},
  journal={IEEE Open Journal of Signal Processing},
  year={2026},
  publisher={IEEE}
}

@article{chen2024vall,
  title={Vall-e 2: Neural codec language models are human parity zero-shot text to speech synthesizers},
  author={Chen, Sanyuan and Liu, Shujie and Zhou, Long and Liu, Yanqing and Tan, Xu and Li, Jinyu and Zhao, Sheng and Qian, Yao and Wei, Furu},
  journal={arXiv preprint arXiv:2406.05370},
  year={2024}
}

@inproceedings{ju2024naturalspeech,
  title={NaturalSpeech 3: Zero-Shot Speech Synthesis with Factorized Codec and Diffusion Models},
  author={Ju, Zeqian and Wang, Yuancheng and Shen, Kai and Tan, Xu and Xin, Detai and Yang, Dongchao and Liu, Eric and Leng, Yichong and Song, Kaitao and Tang, Siliang and others},
  booktitle={International Conference on Machine Learning},
  pages={22605--22623},
  year={2024},
  organization={PMLR}
}

@article{du2024cosyvoice,
  title={Cosyvoice 2: Scalable streaming speech synthesis with large language models},
  author={Du, Zhihao and Wang, Yuxuan and Chen, Qian and Shi, Xian and Lv, Xiang and Zhao, Tianyu and Gao, Zhifu and Yang, Yexin and Gao, Changfeng and Wang, Hui and others},
  journal={arXiv preprint arXiv:2412.10117},
  year={2024}
}

@inproceedings{chen2025f5,
  title={F5-tts: A fairytaler that fakes fluent and faithful speech with flow matching},
  author={Chen, Yushen and Niu, Zhikang and Ma, Ziyang and Deng, Keqi and Wang, Chunhui and JianZhao, JianZhao and Yu, Kai and Chen, Xie},
  booktitle={Proceedings of the 63rd Annual Meeting of the Association for Computational Linguistics (Volume 1: Long Papers)},
  pages={6255--6271},
  year={2025}
}

@article{li2025survey,
  title={A survey on speech deepfake detection},
  author={Li, Menglu and Ahmadiadli, Yasaman and Zhang, Xiao-Ping},
  journal={ACM Computing Surveys},
  volume={57},
  number={7},
  pages={1--38},
  year={2025},
  publisher={ACM New York, NY}
}

@article{chandra2025deepfake,
  title={Deepfake-eval-2024: A multi-modal in-the-wild benchmark of deepfakes circulated in 2024},
  author={Chandra, Nuria Alina and Murtfeldt, Ryan and Qiu, Lin and Karmakar, Arnab and Lee, Hannah and Tanumihardja, Emmanuel and Farhat, Kevin and Caffee, Ben and Paik, Sejin and Lee, Changyeon and others},
  journal={arXiv preprint arXiv:2503.02857},
  year={2025}
}

@article{kwok2025bona,
  title={Bona fide Cross Testing Reveals Weak Spot in Audio Deepfake Detection Systems},
  author={Kwok, Chin Yuen and Yip, Jia Qi and Qiu, Zhen and Chi, Chi Hung and Lam, Kwok Yan},
  journal={arXiv preprint arXiv:2509.09204},
  year={2025}
}

@inproceedings{shi2025benchmarking,
  title={Benchmarking audio deepfake detection robustness in real-world communication scenarios},
  author={Shi, Haohan and Shi, Xiyu and Dogan, Safak and Alzubi, Saif and Huang, Tianjin and Zhang, Yunxiao},
  booktitle={2025 33rd European Signal Processing Conference (EUSIPCO)},
  pages={566--570},
  year={2025},
  organization={IEEE}
}

@inproceedings{wu2020light,
  title={Light Convolutional Neural Network with Feature Genuinization for Detection of Synthetic Speech Attacks},
  author={Wu, Zhenzong and Das, Rohan Kumar and Yang, Jichen and Li, Haizhou},
  booktitle={Proc. Interspeech 2020},
  pages={1101--1105},
  year={2020}
}

@inproceedings{tak2021end,
  title={End-to-end anti-spoofing with rawnet2},
  author={Tak, Hemlata and Patino, Jose and Todisco, Massimiliano and Nautsch, Andreas and Evans, Nicholas and Larcher, Anthony},
  booktitle={ICASSP 2021-2021 IEEE International Conference on Acoustics, Speech and Signal Processing (ICASSP)},
  pages={6369--6373},
  year={2021},
  organization={IEEE}
}

@inproceedings{jung2022aasist,
  title={Aasist: Audio anti-spoofing using integrated spectro-temporal graph attention networks},
  author={Jung, Jee-weon and Heo, Hee-Soo and Tak, Hemlata and Shim, Hye-jin and Chung, Joon Son and Lee, Bong-Jin and Yu, Ha-Jin and Evans, Nicholas},
  booktitle={ICASSP 2022-2022 IEEE international conference on acoustics, speech and signal processing (ICASSP)},
  pages={6367--6371},
  year={2022},
  organization={IEEE}
}

@inproceedings{li2021replay,
  title={Replay and synthetic speech detection with res2net architecture},
  author={Li, Xu and Li, Na and Weng, Chao and Liu, Xunying and Su, Dan and Yu, Dong and Meng, Helen},
  booktitle={ICASSP 2021-2021 IEEE international conference on acoustics, speech and signal processing (ICASSP)},
  pages={6354--6358},
  year={2021},
  organization={IEEE}
}

@inproceedings{zhang2024audio,
  title={Audio deepfake detection with self-supervised xls-r and sls classifier},
  author={Zhang, Qishan and Wen, Shuangbing and Hu, Tao},
  booktitle={Proceedings of the 32nd ACM International Conference on Multimedia},
  pages={6765--6773},
  year={2024}
}

@inproceedings{muller2024harder,
  title={Harder or Different? Understanding Generalization of Audio Deepfake Detection},
  author={M{\"u}ller, Nicolas M and Evans, Nicholas and Tak, Hemlata and Sperl, Philip and B{\"o}ttinger, Konstantin},
  booktitle={Proc. Interspeech 2024},
  pages={2705--2709},
  year={2024}
}

@inproceedings{wang2025mixture,
  title={Mixture of experts fusion for fake audio detection using frozen wav2vec 2.0},
  author={Wang, Zhiyong and Fu, Ruibo and Wen, Zhengqi and Tao, Jianhua and Wang, Xiaopeng and Xie, Yuankun and Qi, Xin and Shi, Shuchen and Lu, Yi and Liu, Yukun and others},
  booktitle={ICASSP 2025-2025 IEEE International Conference on Acoustics, Speech and Signal Processing (ICASSP)},
  pages={1--5},
  year={2025},
  organization={IEEE}
}

@article{li2025frame,
  title={Frame-level Temporal Difference Learning for Partial Deepfake Speech Detection},
  author={Li, Menglu and Zhang, Xiao-Ping and Zhao, Lian},
  journal={IEEE Signal Processing Letters},
  year={2025},
  publisher={IEEE}
}

@inproceedings{tran2025leveraging,
  title={Leveraging SSL Speech Features and Mamba for Enhanced DeepFake Detection},
  author={Tran, Hoan My and Lolive, Damien and Guennec, David and Sini, Aghilas and Delhay, Arnaud and Marteau, Pierre-Fran{\c{c}}ois},
  booktitle={Interspeech 2025},
  pages={5323--5327},
  year={2025}
}

@inproceedings{negroni2025leveraging,
  title={Leveraging mixture of experts for improved speech deepfake detection},
  author={Negroni, Viola and Salvi, Davide and Mezza, Alessandro Ilic and Bestagini, Paolo and Tubaro, Stefano},
  booktitle={ICASSP 2025-2025 IEEE International Conference on Acoustics, Speech and Signal Processing (ICASSP)},
  pages={1--5},
  year={2025},
  organization={IEEE}
}

@article{pan2025molex,
  title={MoLEx: Mixture of LoRA Experts in Speech Self-Supervised Models for Audio Deepfake Detection},
  author={Pan, Zihan and Bhupendra, Sailor Hardik and Wu, Jinyang},
  journal={arXiv preprint arXiv:2509.09175},
  year={2025}
}

@article{cohen2022study,
  title={A study on data augmentation in voice anti-spoofing},
  author={Cohen, Ariel and Rimon, Inbal and Aflalo, Eran and Permuter, Haim H},
  journal={Speech Communication},
  volume={141},
  pages={56--67},
  year={2022},
  publisher={Elsevier}
}

@inproceedings{tak2022rawboost,
  title={Rawboost: A raw data boosting and augmentation method applied to automatic speaker verification anti-spoofing},
  author={Tak, Hemlata and Kamble, Madhu and Patino, Jose and Todisco, Massimiliano and Evans, Nicholas},
  booktitle={ICASSP 2022-2022 IEEE International Conference on Acoustics, Speech and Signal Processing (ICASSP)},
  pages={6382--6386},
  year={2022},
  organization={IEEE}
}

@article{huang2025data,
  title={A Data-Centric Approach to Generalizable Speech Deepfake Detection},
  author={Huang, Wen and Mao, Yuchen and Qian, Yanmin},
  journal={arXiv preprint arXiv:2512.18210},
  year={2025}
}

@article{barrault2023seamless,
  title={Seamless: Multilingual Expressive and Streaming Speech Translation},
  author={Barrault, Lo{\"\i}c and Chung, Yu-An and Meglioli, Mariano Coria and Dale, David and Dong, Ning and Duppenthaler, Mark and Duquenne, Paul-Ambroise and Ellis, Brian and Elsahar, Hady and Haaheim, Justin and others},
  journal={arXiv preprint arXiv:2312.05187},
  year={2023}
}

@article{okabe2018attentive,
  title={Attentive statistics pooling for deep speaker embedding},
  author={Okabe, Koji and Koshinaka, Takafumi and Shinoda, Koichi},
  journal={arXiv preprint arXiv:1803.10963},
  year={2018}
}

@article{india2019self,
  title={Self multi-head attention for speaker recognition},
  author={India, Miquel and Safari, Pooyan and Hernando, Javier},
  journal={arXiv preprint arXiv:1906.09890},
  year={2019}
}

@inproceedings{ko2017study,
  title={A study on data augmentation of reverberant speech for robust speech recognition},
  author={Ko, Tom and Peddinti, Vijayaditya and Povey, Daniel and Seltzer, Michael L and Khudanpur, Sanjeev},
  booktitle={2017 IEEE international conference on acoustics, speech and signal processing (ICASSP)},
  pages={5220--5224},
  year={2017},
  organization={IEEE}
}

@inproceedings{ASVspoof2015,
author = {Wu, Zhizheng and Kinnunen, Tomi and Evans, Nicholas and Yamagishi, Junichi and Hanilçi, Cemal and Sahidullah, Md},
year = {2015},
month = {09},
pages = {},
title = {ASVspoof 2015: the First Automatic Speaker Verification Spoofing and Countermeasures Challenge},
doi = {10.21437/Interspeech.2015-462}
}

@article{wang2020asvspoof2019,
  title   = {{ASVspoof 2019}: A Large-Scale Public Database of Synthesized, Converted and Replayed Speech},
  author  = {Wang, Xin and Yamagishi, Junichi and Todisco, Massimiliano and Delgado, Hector and Nautsch, Andreas and Evans, Nicholas and Sahidullah, Md and Vestman, Ville and Kinnunen, Tomi and Lee, Kong Aik and others},
  journal = {Computer Speech \& Language},
  volume  = {64},
  pages   = {101114},
  year    = {2020},
  doi     = {10.1016/j.csl.2020.101114}
}

@inproceedings{wang2024asvspoof5,
  title={ASVspoof 5: crowdsourced speech data, deepfakes, and adversarial attacks at scale},
  author={Wang, Xin and Delgado, H{\'e}ctor and Tak, Hemlata and Jung, Jee-weon and Shim, Hye-jin and Todisco, Massimiliano and Kukanov, Ivan and Liu, Xuechen and Sahidullah, Md and Kinnunen, Tomi H and others},
  booktitle={Proc. ASVspoof 2024},
  pages={1--8},
  year={2024}
}

@article{muller2022does,
  title={Does audio deepfake detection generalize?},
  author={M{\"u}ller, Nicolas M and Czempin, Pavel and Dieckmann, Franziska and Froghyar, Adam and B{\"o}ttinger, Konstantin},
  journal={arXiv preprint arXiv:2203.16263},
  year={2022}
}

@inproceedings{yamagishi2021asvspoof,
  title={ASVspoof 2021: accelerating progress in spoofed and deepfake speech detection},
  author={Yamagishi, Junichi and Wang, Xin and Todisco, Massimiliano and Sahidullah, Md and Patino, Jose and Nautsch, Andreas and Liu, Xuechen and Lee, Kong Aik and Kinnunen, Tomi and Evans, Nicholas and others},
  booktitle={Proc. ASVSPOOF 2021},
  pages={47--54},
  year={2021}
}

@inproceedings{yi2022add,
  title={Add 2022: the first audio deep synthesis detection challenge},
  author={Yi, Jiangyan and Fu, Ruibo and Tao, Jianhua and Nie, Shuai and Ma, Haoxin and Wang, Chenglong and Wang, Tao and Tian, Zhengkun and Bai, Ye and Fan, Cunhang and others},
  booktitle={ICASSP 2022-2022 IEEE International Conference on Acoustics, Speech and Signal Processing (ICASSP)},
  pages={9216--9220},
  year={2022},
  organization={IEEE}
}

@article{yi2023add,
  title={Add 2023: the second audio deepfake detection challenge},
  author={Yi, Jiangyan and Tao, Jianhua and Fu, Ruibo and Yan, Xinrui and Wang, Chenglong and Wang, Tao and Zhang, Chu Yuan and Zhang, Xiaohui and Zhao, Yan and Ren, Yong and others},
  journal={arXiv preprint arXiv:2305.13774},
  year={2023}
}

@inproceedings{for2019,
  title={FoR: A Dataset for Synthetic Speech Detection},
  author={Reimao, Ricardo and Tzerpos, Vassilios},
  booktitle={2019 International Conference on Speech Technology and Human-Computer Dialogue (SpeD)},
  pages={1--10},
  year={2019},
  organization={IEEE}
}

@article{jung2025spoofceleb,
  title={Spoofceleb: Speech deepfake detection and sasv in the wild},
  author={Jung, Jee-weon and Wu, Yihan and Wang, Xin and Kim, Ji-Hoon and Maiti, Soumi and Matsunaga, Yuta and Shim, Hye-jin and Tian, Jinchuan and Evans, Nicholas and Chung, Joon Son and others},
  journal={IEEE Open Journal of Signal Processing},
  year={2025},
  publisher={IEEE}
}

@inproceedings{muller2025replay,
  title={Replay Attacks Against Audio Deepfake Detection},
  author={M{\"u}ller, Nicolas and Kawa, Piotr and Choong, Wei-Herng and Stan, Adriana and Bukkapatnam, Aditya Tirumala and Pizzi, Karla and Wagner, Alexander and Sperl, Philip},
  booktitle={Proc. Interspeech 2025},
  pages={2245--2249},
  year={2025}
}

@inproceedings{du2024dfadd,
  title={Dfadd: The diffusion and flow-matching based audio deepfake dataset},
  author={Du, Jiawei and Lin, I-Ming and Chiu, I-Hsiang and Chen, Xuanjun and Wu, Haibin and Ren, Wenze and Tsao, Yu and Lee, Hung-Yi and Jang, Jyh-Shing Roger},
  booktitle={2024 IEEE Spoken Language Technology Workshop (SLT)},
  pages={921--928},
  year={2024},
  organization={IEEE}
}

@inproceedings{muller2024mlaad,
  title={Mlaad: The multi-language audio anti-spoofing dataset},
  author={M{\"u}ller, Nicolas M and Kawa, Piotr and Choong, Wei Herng and Casanova, Edresson and G{\"o}lge, Eren and M{\"u}ller, Thorsten and Syga, Piotr and Sperl, Philip and B{\"o}ttinger, Konstantin},
  booktitle={2024 International Joint Conference on Neural Networks (IJCNN)},
  pages={1--7},
  year={2024},
  organization={IEEE}
}

@inproceedings{sun2023ai,
  title={Ai-synthesized voice detection using neural vocoder artifacts},
  author={Sun, Chengzhe and Jia, Shan and Hou, Shuwei and Lyu, Siwei},
  booktitle={Proceedings of the IEEE/CVF Conference on Computer Vision and Pattern Recognition},
  pages={904--912},
  year={2023}
}

@inproceedings{huang2025speechfake,
  title={SpeechFake: A large-scale multilingual speech deepfake dataset incorporating cutting-edge generation methods},
  author={Huang, Wen and Gu, Yanmei and Wang, Zhiming and Zhu, Huijia and Qian, Yanmin},
  booktitle={Proceedings of the 63rd Annual Meeting of the Association for Computational Linguistics (Volume 1: Long Papers)},
  pages={9985--9998},
  year={2025}
}

@article{frank2021wavefake,
  title={Wavefake: A data set to facilitate audio deepfake detection},
  author={Frank, Joel and Sch{\"o}nherr, Lea},
  journal={arXiv preprint arXiv:2111.02813},
  year={2021}
}

@article{xie2025codecfake,
  title={The codecfake dataset and countermeasures for the universally detection of deepfake audio},
  author={Xie, Yuankun and Lu, Yi and Fu, Ruibo and Wen, Zhengqi and Wang, Zhiyong and Tao, Jianhua and Qi, Xin and Wang, Xiaopeng and Liu, Yukun and Cheng, Haonan and others},
  journal={IEEE Transactions on Audio, Speech and Language Processing},
  volume={33},
  pages={386--400},
  year={2025},
  publisher={IEEE}
}

@article{li2410sonar,
  title={Sonar: a synthetic AI-audio detection framework and benchmark (2024)},
  author={Li, X and Chen, PY and Wei, W},
  journal={arXiv preprint arXiv:2410.04324},
  year={2024}
}

@inproceedings{panayotov2015librispeech,
  title={Librispeech: an asr corpus based on public domain audio books},
  author={Panayotov, Vassil and Chen, Guoguo and Povey, Daniel and Khudanpur, Sanjeev},
  booktitle={2015 IEEE international conference on acoustics, speech and signal processing (ICASSP)},
  pages={5206--5210},
  year={2015},
  organization={IEEE}
}

@article{shi2020aishell,
  title={Aishell-3: A multi-speaker mandarin tts corpus and the baselines},
  author={Shi, Yao and Bu, Hui and Xu, Xin and Zhang, Shaoji and Li, Ming},
  journal={arXiv preprint arXiv:2010.11567},
  year={2020}
}

@article{chen2021gigaspeech,
  title={Gigaspeech: An evolving, multi-domain asr corpus with 10,000 hours of transcribed audio},
  author={Chen, Guoguo and Chai, Shuzhou and Wang, Guanbo and Du, Jiayu and Zhang, Wei-Qiang and Weng, Chao and Su, Dan and Povey, Daniel and Trmal, Jan and Zhang, Junbo and others},
  journal={arXiv preprint arXiv:2106.06909},
  year={2021}
}

@inproceedings{fan2020cn,
  title={Cn-celeb: a challenging chinese speaker recognition dataset},
  author={Fan, Yue and Kang, JW and Li, LT and Li, KC and Chen, HL and Cheng, ST and Zhang, PY and Zhou, ZY and Cai, YQ and Wang, Dong},
  booktitle={ICASSP 2020-2020 IEEE International Conference on Acoustics, Speech and Signal Processing (ICASSP)},
  pages={7604--7608},
  year={2020},
  organization={IEEE}
}

@inproceedings{ardila2020common,
  title={Common voice: A massively-multilingual speech corpus},
  author={Ardila, Rosana and Branson, Megan and Davis, Kelly and Kohler, Michael and Meyer, Josh and Henretty, Michael and Morais, Reuben and Saunders, Lindsay and Tyers, Francis and Weber, Gregor},
  booktitle={Proceedings of the twelfth language resources and evaluation conference},
  pages={4218--4222},
  year={2020}
}

@article{snyder2015musan,
  title={Musan: A music, speech, and noise corpus},
  author={Snyder, David and Chen, Guoguo and Povey, Daniel},
  journal={arXiv preprint arXiv:1510.08484},
  year={2015}
}

@article{ge2025post,
  title={Post-training for deepfake speech detection},
  author={Ge, Wanying and Wang, Xin and Liu, Xuechen and Yamagishi, Junichi},
  journal={arXiv preprint arXiv:2506.21090},
  year={2025}
}

@misc{modulate_velma2_synthetic_voice_2026,
  author       = {{Modulate}},
  title        = {{Modulate-VELMA-2-Synthetic-Voice}},
  year         = {2026},
  howpublished = {\url{https://www.modulate.ai/benchmarks}}
}

@misc{resemble_detect_3b_omni_2026,
  author       = {{Resemble AI}},
  title        = {{Resemble-Detect-3B-Omni}},
  year         = {2025},
  howpublished = {\url{https://www.resemble.ai/}}
}

@misc{hiya_authenticity_verification_multi_v1_2026,
  author       = {{Hiya}},
  title        = {{Hiya-Authenticity-Verification-Multi-v1}},
  year         = {2026},
  howpublished = {\url{https://www.hiya.com/}}
}

@misc{dlmsl_speaksure_v01_2026,
  author       = {{DLMSL}},
  title        = {{DLMSL-SpeakSure-v0.1}},
  year         = {2025},
  howpublished = {\url{https://dlmsl.csie.ncu.edu.tw/}}
}

@misc{df_raptor_2026,
  author       = {{Idiap}},
  title        = {{DF-Raptor}},
  year         = {2026},
  howpublished = {\url{https://www.idiap.ch/en/}}
}

@misc{whispeak_2026,
  author       = {{Whispeak}},
  title        = {{Whispeak}},
  year         = {2025},
  howpublished = {\url{https://whispeak.io/}}
}

@misc{kulkarni2026compactsslbackbonesmatter,
      title={Do Compact SSL Backbones Matter for Audio Deepfake Detection? A Controlled Study with RAPTOR}, 
      author={Ajinkya Kulkarni and Sandipana Dowerah and Atharva Kulkarni and Tanel Alumäe and Mathew Magimai Doss},
      year={2026},
      eprint={2603.06164},
      archivePrefix={arXiv},
      primaryClass={cs.SD} 
}

@misc{resemble_detect_2026,
  author       = {{Resemble AI}},
  title        = {{Resemble Detect}},
  year         = {2025},
  howpublished = {\url{https://www.resemble.ai/}}
}

@article{truong2024temporal,
  title={Temporal-channel modeling in multi-head self-attention for synthetic speech detection},
  author={Truong, Duc-Tuan and Tao, Ruijie and Nguyen, Tuan and Luong, Hieu-Thi and Lee, Kong Aik and Chng, Eng Siong},
  journal={arXiv preprint arXiv:2406.17376},
  year={2024}
}

@article{kheir2025bicrossmamba,
  title={BiCrossMamba-ST: speech deepfake detection with bidirectional mamba spectro-temporal cross-attention},
  author={Kheir, Yassine El and Polzehl, Tim and M{\"o}ller, Sebastian},
  journal={arXiv preprint arXiv:2505.13930},
  year={2025}
}

@article{liu2025nes2net,
  title={Nes2net: A lightweight nested architecture for foundation model driven speech anti-spoofing},
  author={Liu, Tianchi and Truong, Duc-Tuan and Das, Rohan Kumar and Lee, Kong Aik and Li, Haizhou},
  journal={IEEE Transactions on Information Forensics and Security},
  volume={20},
  pages={12005--12018},
  year={2025},
  publisher={IEEE}
}

@article{tak2022automatic,
  title={Automatic speaker verification spoofing and deepfake detection using wav2vec 2.0 and data augmentation},
  author={Tak, Hemlata and Todisco, Massimiliano and Wang, Xin and Jung, Jee-weon and Yamagishi, Junichi and Evans, Nicholas},
  journal={arXiv preprint arXiv:2202.12233},
  year={2022}
}

@article{xiao2025xlsr,
  title={XLSR-Mamba: A dual-column bidirectional state space model for spoofing attack detection},
  author={Xiao, Yang and Das, Rohan Kumar},
  journal={IEEE Signal Processing Letters},
  year={2025},
  publisher={IEEE}
}

@article{kawa2023improved,
  title={Improved deepfake detection using whisper features},
  author={Kawa, Piotr and Plata, Marcin and Czuba, Micha{\'L} and Syga, Piotr and others},
  journal={arXiv preprint arXiv:2306.01428},
  year={2023}
}

@inproceedings{kulkarni2024exploring,
  title={Exploring generalization to unseen audio data for spoofing: Insights from SSL models},
  author={Kulkarni, Atharva and Tran, Hoan My and Kulkarni, Ajinkya and Dowerah, Sandipana and Lolive, Damien and Doss, Mathew Magimai},
  booktitle={ASVSpoof workshop 2024},
  year={2024}
}

@article{tak2021end_2,
  title={End-to-end spectro-temporal graph attention networks for speaker verification anti-spoofing and speech deepfake detection},
  author={Tak, Hemlata and Jung, Jee-weon and Patino, Jose and Kamble, Madhu and Todisco, Massimiliano and Evans, Nicholas},
  journal={arXiv preprint arXiv:2107.12710},
  year={2021}
}

@inproceedings{xiao2025rawtfnet,
  title={RawTFNet: A Lightweight CNN Architecture for Speech Anti-spoofing},
  author={Xiao, Yang and Dang, Ting and Das, Rohan Kumar},
  booktitle={2025 Asia Pacific Signal and Information Processing Association Annual Summit and Conference (APSIPA ASC)},
  pages={1997--2001},
  year={2025},
  organization={IEEE}
}

@misc{syntra_detector_2026,
  author       = {{Syntra}},
  title        = {{Syntra Detector}},
  year         = {2025},
  howpublished = {\url{https://syntra.io/}}
}

@article{chen2022wavlm,
  title={Wavlm: Large-scale self-supervised pre-training for full stack speech processing},
  author={Chen, Sanyuan and Wang, Chengyi and Chen, Zhengyang and Wu, Yu and Liu, Shujie and Chen, Zhuo and Li, Jinyu and Kanda, Naoyuki and Yoshioka, Takuya and Xiao, Xiong and others},
  journal={IEEE Journal of Selected Topics in Signal Processing},
  volume={16},
  number={6},
  pages={1505--1518},
  year={2022},
  publisher={IEEE}
}

@article{babu2021xls,
  title={XLS-R: Self-supervised cross-lingual speech representation learning at scale},
  author={Babu, Arun and Wang, Changhan and Tjandra, Andros and Lakhotia, Kushal and Xu, Qiantong and Goyal, Naman and Singh, Kritika and Von Platen, Patrick and Saraf, Yatharth and Pino, Juan and others},
  journal={arXiv preprint arXiv:2111.09296},
  year={2021}
}

@article{baevski2020wav2vec,
  title={wav2vec 2.0: A framework for self-supervised learning of speech representations},
  author={Baevski, Alexei and Zhou, Yuhao and Mohamed, Abdelrahman and Auli, Michael},
  journal={Advances in neural information processing systems},
  volume={33},
  pages={12449--12460},
  year={2020}
}

@article{defossez2022high,
  title={High fidelity neural audio compression},
  author={D{\'e}fossez, Alexandre and Copet, Jade and Synnaeve, Gabriel and Adi, Yossi},
  journal={arXiv preprint arXiv:2210.13438},
  year={2022}
}

@inproceedings{gulati2020conformer,
  title={Conformer: Convolution-augmented Transformer for Speech Recognition},
  author={Gulati, Anmol and Qin, James and Chiu, Chung-Cheng and Parmar, Niki and Zhang, Yu and Yu, Jiahui and Han, Wei and Wang, Shibo and Zhang, Zhengdong and Wu, Yonghui and others},
  booktitle={Proc. Interspeech 2020},
  pages={5036--5040},
  year={2020}
}

@inproceedings{ye2025data,
  title={Data mixing laws: Optimizing data mixtures by predicting language modeling performance},
  author={Ye, Jiasheng and Liu, Peiju and Sun, Tianxiang and Zhan, Jun and Zhou, Yunhua and Qiu, Xipeng},
  booktitle={International Conference on Learning Representations},
  volume={2025},
  pages={82263--82287},
  year={2025}
}

@article{xie2023doremi,
  title={Doremi: Optimizing data mixtures speeds up language model pretraining},
  author={Xie, Sang Michael and Pham, Hieu and Dong, Xuanyi and Du, Nan and Liu, Hanxiao and Lu, Yifeng and Liang, Percy S and Le, Quoc V and Ma, Tengyu and Yu, Adams Wei},
  journal={Advances in Neural Information Processing Systems},
  volume={36},
  pages={69798--69818},
  year={2023}
}

@inproceedings{
    chungunimax,
    title={UniMax: Fairer and More Effective Language Sampling for Large-Scale Multilingual Pretraining},
    author={Hyung Won Chung and Xavier Garcia and Adam Roberts and Yi Tay and Orhan Firat and Sharan Narang and Noah Constant},
    booktitle={The Eleventh International Conference on Learning Representations },
    year={2023},
    url={https://openreview.net/forum?id=kXwdL1cWOAi}
}

@inproceedings{huang2025generalizable,
  title={Generalizable audio deepfake detection via latent space refinement and augmentation},
  author={Huang, Wen and Gu, Yanmei and Wang, Zhiming and Zhu, Huijia and Qian, Yanmin},
  booktitle={ICASSP 2025-2025 IEEE International Conference on Acoustics, Speech and Signal Processing (ICASSP)},
  pages={1--5},
  year={2025},
  organization={IEEE}
}

@inproceedings{laakkonen2026generalizable,
  title={Generalizable speech deepfake detection via meta-learned lora},
  author={Laakkonen, Janne and Kukanov, Ivan and Hautam{\"a}ki, Ville},
  booktitle={ICASSP 2026-2026 IEEE International Conference on Acoustics, Speech and Signal Processing (ICASSP)},
  pages={19092--19096},
  year={2026},
  organization={IEEE}
}

@inproceedings{huang2026generalizable,
  title={Generalizable speech deepfake detection via information bottleneck enhanced adversarial alignment},
  author={Huang, Pu and Wang, Shouguang and Yao, Siya and Zhou, Mengchu},
  booktitle={ICASSP 2026-2026 IEEE International Conference on Acoustics, Speech and Signal Processing (ICASSP)},
  pages={19087--19091},
  year={2026},
  organization={IEEE}
}

@inproceedings{yang2025generalizable,
  title={Generalizable Audio Deepfake Detection via Hierarchical Structure Learning and Feature Whitening in Poincar{\'e} sphere},
  author={Yang, Mingru and Gu, Yanmei and He, Qianhua and Li, Yanxiong and Zhang, Peirong and Chen, Yongqiang and Wang, Zhiming and Zhu, Huijia and Liu, Jian and Wang, Weiqiang},
  booktitle={Proc. Interspeech 2025},
  pages={2255--2259},
  year={2025}
}

\clearpage
\appendix
\section{Additional Leaderboard Performances}

Tables~\ref{tab:leaderboard_acc} and~\ref{tab:leaderboard_f1} report additional ACC and F1-score comparisons on the Speech-DF-Arena. Teffic-Audio achieves the best pooled ACC and pooled F1-score among the listed systems. Together with the EER results in the main text, these results show a consistent performance advantage across different measurements, further supporting the stability of Teffic-Audio in the comprehensive cross-dataset evaluation.

\begin{table*}[htbp]
    \centering
    \caption{ACC performance on Speech-DF-Arena.}
    \label{tab:leaderboard_acc}
    \tiny
    \setlength{\syscellheight}{2\baselineskip}
    \setlength{\tabcolsep}{2.2pt}
    \resizebox{\linewidth}{!}{%
    \begin{tabular}{p{2.75cm}|*{4}{c}|*{14}{c}}
    \toprule
    \rowcolor{headergray}
     & & & & & &
    \multicolumn{4}{c}{\textbf{ASVspoof}} &
     & &
    \multicolumn{4}{c@{}}{\textbf{ADD}} &
     & & \\
    \noalign{\global\aboverulesep=0pt\global\belowrulesep=0pt}
    \cmidrule(lr){7-10}\cmidrule(lr){13-16}
    \noalign{\global\aboverulesep=.4ex\global\belowrulesep=.65ex}
    \rowcolor{headergray}
    \multirow{-2}{*}{\textbf{System}} &
    \multirow{-2}{*}{\textbf{Licence}} &
    \multirow{-2}{*}{\textbf{Params}} &
    \multirow{-2}{*}{\textbf{Pooled}} &
    \multirow{-2}{*}{\textbf{Avg.}} &
    \multirow{-2}{*}{\textbf{ITW}} &
    \multicolumn{1}{c}{\textbf{19}} &
    \multicolumn{1}{c}{\textbf{21-LA}} &
    \multicolumn{1}{c}{\textbf{21-DF}} &
    \multicolumn{1}{c}{\textbf{24-E}} &
    \multicolumn{1}{c}{\multirow{-2}{*}{\textbf{FoR}}} &
    \multicolumn{1}{c}{\multirow{-2}{*}{\textbf{CF}}} &
    \multicolumn{1}{c}{\textbf{22-T1}} &
    \multicolumn{1}{c}{\textbf{22-T3}} &
    \multicolumn{1}{c}{\textbf{23-R1}} &
    \multicolumn{1}{c}{\textbf{23-R2}} &
    \multirow{-2}{*}{\textbf{DFADD}} &
    \multirow{-2}{*}{\textbf{LSV}} &
    \multirow{-2}{*}{\textbf{SONAR}} \\
    \midrule

    \rowcolor{oursrow}
    \sysone{Teffic-Audio} & Proprietary & 590.0 & \textbf{98.546} & 98.759 & 98.682 & 99.761 & 97.961 & 99.738 & 98.595 & 99.006 & \textbf{99.251} & 92.969 & \textbf{99.162} & \textbf{99.645} & 98.163 & \textbf{99.973} & \textbf{99.995} & 99.721 \\
    \midrule

    \systwo{Modulate-VELMA-2-Synthetic-}{Voice~\citep{modulate_velma2_synthetic_voice_2026}}
    & Proprietary & 316.0 & 98.414 & \textbf{98.897} & 98.726 & 99.702 & 98.670 & 99.667 & \textbf{99.616} & \textbf{99.890} & 98.463 & \textbf{94.940} & 98.825 & 98.958 & \textbf{98.257} & \textbf{99.973} & 99.730 & 99.139 \\

    \syscite{Resemble-Detect-3B-Omni}{resemble_detect_3b_omni_2026}
    & Proprietary & 3000.0 & 97.901 & 97.430 & 98.656 & 97.469 & 96.984 & 99.421 & 99.547 & 97.637 & 97.312 & 90.043 & 98.453 & 96.611 & 92.641 & 99.947 & \textbf{99.995} & 98.835 \\

    \systwo{Hiya-Authenticity-Verification}{Multi-v1~\citep{hiya_authenticity_verification_multi_v1_2026}}
    & Proprietary & 1000.0 & 97.680 & 97.880 & 99.330 & 99.700 & 98.990 & 98.680 & 99.210 & 99.800 & 94.270 & 87.900 & 98.810 & 98.020 & 96.000 & 99.950 & 99.990 & 99.540 \\

    \syscite{DLMSL-SpeakSure-v0.1}{dlmsl_speaksure_v01_2026}
    & Proprietary & 658.6 & 93.858 & 96.043 & 98.726 & \textbf{99.959} & \textbf{99.920} & \textbf{99.986} & 87.105 & 99.757 & 93.173 & 84.859 & 97.627 & 94.347 & 91.721 & 99.840 & 99.870 & \textbf{99.899} \\

    \sysone{DF-Raptor~\citep{df_raptor_2026}}
    & Proprietary & 100.0 & 92.330 & 92.080 & 96.523 & 95.870 & 93.890 & 98.025 & 93.140 & 96.952 & 84.447 & 69.436 & 96.215 & 92.974 & 93.335 & 97.870 & 92.311 & 88.222 \\

    \sysone{Whispeak~\citep{whispeak_2026}}
    & Proprietary & 98.9 & 91.953 & 96.947 & 98.729 & 99.604 & 96.420 & 96.766 & 90.075 & 99.006 & 99.144 & 88.057 & 97.689 & 97.381 & 94.994 & \textbf{99.973} & 99.946 & 99.443 \\

    \syscite{DF\_Arena\_1B\_V\_1}{kulkarni2026compactsslbackbonesmatter}
    & Open & 1000.0 & 90.476 & 94.079 & 99.097 & 98.864 & 95.342 & 98.251 & 82.750 & 97.107 & 91.631 & 77.791 & 97.798 & 94.917 & 88.457 & \textbf{99.973} & 99.843 & 98.886 \\

    \sysone{Syntra Detector~\citep{syntra_detector_2026}}
    & Proprietary & 584.0 & 89.236 & 93.893 & 96.019 & 98.516 & 85.946 & 97.972 & 84.040 & 99.492 & 98.816 & 76.417 & 97.013 & 96.348 & 90.397 & \textbf{99.973} & 98.837 & 99.367 \\

    \syscite{DF\_Arena\_500M\_V\_1}{kulkarni2026compactsslbackbonesmatter}
    & Open & 500.0 & 89.110 & 94.190 & 98.240 & 98.910 & 95.770 & 96.700 & 87.610 & 97.730 & 93.650 & 76.020 & 97.230 & 92.530 & 87.700 & 99.970 & 99.870 & 98.130 \\

    \sysone{MoLEx~\citep{pan2025molex}}
    & Proprietary & 376.4 & 87.595 & 90.483 & \textbf{99.969} & 99.719 & 93.673 & 98.114 & 84.130 & 99.845 & 67.600 & 68.062 & 96.338 & 88.890 & 80.978 & 93.582 & 99.692 & 99.063 \\

    \syscite{Resemble Detect}{resemble_detect_2026}
    & Proprietary & 2112.0 & 87.253 & 89.170 & 96.051 & 98.679 & 98.353 & 96.207 & 83.709 & 98.653 & 66.959 & 71.781 & 93.886 & 78.928 & 71.723 & \textbf{99.973} & 98.366 & 97.036 \\

    \sysone{DF\_Arena\_100M\_V\_1}
    & Closed & 100.0 & 86.079 & 91.605 & 97.778 & 98.467 & 92.389 & 94.499 & 78.601 & 92.602 & 91.252 & 72.937 & 94.578 & 91.097 & 82.960 & \textbf{99.973} & 99.789 & 95.542 \\

    \sysone{DF\_Arena\_100M\_V\_0}
    & Closed & 100.0 & 84.060 & 89.070 & 95.710 & 95.830 & 89.940 & 91.170 & 78.720 & 93.840 & 89.610 & 68.050 & 93.310 & 87.310 & 80.990 & 99.890 & 99.580 & 94.300 \\

    \sysone{XLSR+SLS~\citep{zhang2024audio}}
    & Open & 340.0 & 83.921 & 85.984 & 92.542 & 99.770 & 97.129 & 98.089 & 81.236 & 94.898 & 66.563 & 66.051 & 84.258 & 80.624 & 78.905 & 92.437 & 98.026 & 75.304 \\

    \sysone{TCM~\citep{truong2024temporal}}
    & Open & 319.0 & 83.309 & 84.153 & 92.209 & 99.815 & 97.006 & 97.854 & 81.150 & 89.289 & 63.992 & 62.599 & 79.060 & 76.574 & 77.256 & 91.079 & 97.647 & 73.455 \\

    \syscite{BiCrossMamba-ST}{kheir2025bicrossmamba}
    & Open & 318.2 & 82.846 & 84.219 & 92.061 & 99.291 & 96.172 & 97.653 & 86.331 & 93.132 & 62.298 & 69.556 & 81.316 & 70.557 & 70.070 & 91.505 & 97.874 & 72.619 \\

    \sysone{Nes2NetX~\citep{liu2025nes2net}}
    & Open & 317.9 & 82.634 & 83.814 & 92.253 & 99.879 & 97.828 & 98.507 & 77.940 & 93.706 & 60.659 & 65.530 & 73.443 & 78.864 & 81.549 & 88.815 & 97.117 & 68.490 \\

    \syscite{Wav2Vec2 AASIST}{tak2022automatic}
    & Open & 317.8 & 80.393 & 81.867 & 88.807 & 99.775 & 99.176 & 93.371 & 83.753 & 92.557 & 56.634 & 68.949 & 83.472 & 72.257 & 78.072 & 88.096 & 88.787 & 53.901 \\

    \syscite{XLSR Mamba}{xiao2025xlsr}
    & Open & 319.0 & 79.409 & 85.355 & 93.294 & 99.577 & 99.068 & 98.115 & 85.599 & 93.308 & 64.736 & 65.771 & 80.631 & 78.152 & 79.850 & 89.348 & 97.761 & 75.709 \\

    \syscite{Whisper Mesonet}{kawa2023improved}
    & Open & 7.6 & 76.449 & 71.649 & 73.272 & 94.166 & 84.178 & 97.889 & 77.451 & 52.275 & 65.338 & 61.616 & 75.991 & 58.751 & 55.437 & 75.899 & 84.659 & 41.363 \\

    \syscite{Wav2Vec2 ECAPA}{kulkarni2024exploring}
    & Open & 324.0 & 71.005 & 62.084 & 69.307 & 70.303 & 73.395 & 77.561 & 81.344 & 37.699 & 59.026 & 53.566 & 78.128 & 64.713 & 63.299 & 24.980 & 71.477 & 35.461 \\

    \sysone{AASIST~\citep{jung2022aasist}}
    & Open & 0.3 & 67.229 & 65.597 & 56.994 & 99.169 & 88.539 & 78.930 & 64.466 & 78.379 & 48.941 & 52.080 & 66.813 & 52.244 & 67.530 & 58.109 & 61.973 & 42.553 \\

    \syscite{WavLM ECAPA}{kulkarni2024exploring}
    & Open & 102.0 & 66.536 & 71.010 & 65.348 & 99.238 & 93.324 & 84.056 & 74.008 & 76.656 & 53.816 & 55.830 & 60.588 & 71.001 & 68.064 & 70.493 & 68.113 & 58.080 \\

    \sysone{RawGatST~\citep{tak2021end_2}}
    & Open & 0.4 & 66.447 & 65.181 & 47.465 & 98.939 & 89.747 & 76.740 & 59.709 & 46.930 & 50.002 & 57.098 & 67.698 & 62.135 & 72.662 & 76.272 & 56.624 & 49.240 \\

    \sysone{RawTFNet~\citep{xiao2025rawtfnet}}
    & Open & 0.2 & 60.062 & 67.178 & 61.276 & 98.109 & 94.961 & 83.180 & 55.260 & 63.450 & 48.223 & 56.151 & 63.192 & 61.275 & 69.460 & 77.470 & 70.412 & 45.187 \\

    \syscite{Hubert ECAPA}{kulkarni2024exploring}
    & Open & 102.0 & 56.844 & 66.155 & 61.339 & 98.946 & 87.445 & 86.208 & 68.604 & 66.232 & 53.777 & 52.339 & 60.915 & 50.432 & 56.049 & 65.406 & 67.961 & 59.828 \\

    \sysone{Rawnet2~\citep{tak2021end}}
    & Open & 17.6 & 54.004 & 52.434 & 50.817 & 66.960 & 59.929 & 59.331 & 58.781 & 34.342 & 49.781 & 50.641 & 52.372 & 44.360 & 35.447 & 60.905 & 51.826 & 57.016 \\

    \bottomrule
    \end{tabular}
    }%
\end{table*}

\makeatletter
\setlength{\@dblfptop}{0pt}
\makeatother
\begin{table*}[!t]
    \centering
    \caption{F1-score performance on Speech-DF-Arena}
    \label{tab:leaderboard_f1}
    \tiny
    \setlength{\syscellheight}{2\baselineskip}
    \setlength{\tabcolsep}{2.2pt}
    \resizebox{\linewidth}{!}{%
    \begin{tabular}{p{2.75cm}|*{4}{c}|*{14}{c}}
    \toprule
    \rowcolor{headergray}
     & & & & & &
    \multicolumn{4}{c}{\textbf{ASVspoof}} &
     & &
    \multicolumn{4}{c@{}}{\textbf{ADD}} &
     & & \\
    \noalign{\global\aboverulesep=0pt\global\belowrulesep=0pt}
    \cmidrule(lr){7-10}\cmidrule(lr){13-16}
    \noalign{\global\aboverulesep=.4ex\global\belowrulesep=.65ex}
    \rowcolor{headergray}
    \multirow{-2}{*}{\textbf{System}} &
    \multirow{-2}{*}{\textbf{Licence}} &
    \multirow{-2}{*}{\textbf{Params}} &
    \multirow{-2}{*}{\textbf{Pooled}} &
    \multirow{-2}{*}{\textbf{Avg.}} &
    \multirow{-2}{*}{\textbf{ITW}} &
    \multicolumn{1}{c}{\textbf{19}} &
    \multicolumn{1}{c}{\textbf{21-LA}} &
    \multicolumn{1}{c}{\textbf{21-DF}} &
    \multicolumn{1}{c}{\textbf{24-E}} &
    \multicolumn{1}{c}{\multirow{-2}{*}{\textbf{FoR}}} &
    \multicolumn{1}{c}{\multirow{-2}{*}{\textbf{CF}}} &
    \multicolumn{1}{c}{\textbf{22-T1}} &
    \multicolumn{1}{c}{\textbf{22-T3}} &
    \multicolumn{1}{c}{\textbf{23-R1}} &
    \multicolumn{1}{c}{\textbf{23-R2}} &
    \multirow{-2}{*}{\textbf{DFADD}} &
    \multirow{-2}{*}{\textbf{LSV}} &
    \multirow{-2}{*}{\textbf{SONAR}} \\
    \midrule

    \rowcolor{oursrow}
    \sysone{Teffic-Audio} & Proprietary & 590.0 & \textbf{0.969} & 0.973 & 0.989 & 0.989 & 0.906 & 0.955 & 0.966 & 0.990 & \textbf{0.988} & 0.884 & \textbf{0.975} & \textbf{0.998} & 0.987 & \textbf{0.999} & \textbf{1.000} & 0.998 \\
    \midrule

    \systwo{Modulate-VELMA-2-Synthetic-}{Voice~\citep{modulate_velma2_synthetic_voice_2026}}
    & Proprietary & 316.0 & 0.966 & 0.976 & 0.990 & 0.986 & 0.937 & 0.943 & \textbf{0.991} & 0.999 & 0.975 & \textbf{0.915} & 0.965 & 0.993 & \textbf{0.988} & 0.999 & 0.991 & 0.993 \\

    \syscite{Resemble-Detect-3B-Omni}{resemble_detect_3b_omni_2026}
    & Proprietary & 3000.0 & 0.955 & 0.949 & 0.989 & 0.888 & 0.865 & 0.905 & 0.989 & 0.976 & 0.957 & 0.838 & 0.954 & 0.976 & 0.949 & 0.999 & \textbf{1.000} & 0.990 \\

    \systwo{Hiya-Authenticity-Verification}{Multi-v1~\citep{hiya_authenticity_verification_multi_v1_2026}}
    & Proprietary & 1000.0 & 0.950 & 0.954 & 0.990 & 0.990 & 0.950 & 0.810 & 0.980 & \textbf{1.000} & 0.910 & 0.810 & 0.960 & 0.990 & 0.970 & \textbf{1.000} & \textbf{1.000} & \textbf{1.000} \\

    \syscite{DLMSL-SpeakSure-v0.1}{dlmsl_speaksure_v01_2026}
    & Proprietary & 658.6 & 0.874 & 0.938 & 0.990 & 0.998 & \textbf{0.996} & \textbf{0.998} & 0.733 & 0.998 & 0.893 & 0.763 & 0.931 & 0.960 & 0.942 & 0.996 & 0.995 & 0.999 \\

    \sysone{DF-Raptor~\citep{df_raptor_2026}}
    & Proprietary & 100.0 & 0.840 & 0.836 & 0.972 & 0.807 & 0.691 & 0.605 & 0.840 & 0.969 & 0.785 & 0.612 & 0.886 & 0.950 & 0.955 & 0.944 & 0.787 & 0.907 \\

    \sysone{Whispeak~\citep{whispeak_2026}}
    & Proprietary & 98.9 & 0.834 & 0.925 & 0.990 & 0.981 & 0.843 & 0.625 & 0.787 & 0.990 & 0.994 & 0.809 & 0.933 & 0.982 & 0.966 & 0.999 & 0.998 & 0.995 \\

    \syscite{DF\_Arena\_1B\_V\_1}{kulkarni2026compactsslbackbonesmatter}
    & Open & 1000.0 & 0.812 & 0.886 & 0.993 & 0.947 & 0.804 & 0.758 & 0.662 & 0.971 & 0.870 & 0.668 & 0.936 & 0.964 & 0.919 & 0.999 & 0.995 & 0.990 \\

    \sysone{Syntra Detector~\citep{syntra_detector_2026}}
    & Proprietary & 584.0 & 0.790 & 0.870 & 0.968 & 0.932 & 0.550 & 0.729 & 0.682 & 0.995 & 0.981 & 0.650 & 0.914 & 0.974 & 0.933 & 0.999 & 0.960 & 0.994 \\

    \syscite{DF\_Arena\_500M\_V\_1}{kulkarni2026compactsslbackbonesmatter}
    & Open & 500.0 & 0.780 & 0.884 & 0.986 & 0.949 & 0.819 & 0.620 & 0.742 & 0.977 & 0.900 & 0.645 & 0.920 & 0.947 & 0.913 & 0.999 & 0.996 & 0.984 \\

    \sysone{MoLEx~\citep{pan2025molex}}
    & Proprietary & 376.4 & 0.763 & 0.836 & \textbf{1.000} & 0.987 & 0.748 & 0.743 & 0.684 & 0.998 & 0.559 & 0.550 & 0.896 & 0.920 & 0.863 & 0.854 & 0.989 & 0.992 \\

    \syscite{Resemble Detect}{resemble_detect_2026}
    & Proprietary & 2112.0 & 0.759 & 0.825 & 0.968 & 0.939 & 0.923 & 0.586 & 0.677 & 0.987 & 0.552 & 0.593 & 0.834 & 0.843 & 0.789 & 0.999 & 0.945 & 0.974 \\

    \sysone{DF\_Arena\_100M\_V\_1}
    & Closed & 100.0 & 0.738 & 0.837 & 0.982 & 0.930 & 0.708 & 0.489 & 0.599 & 0.926 & 0.864 & 0.607 & 0.851 & 0.936 & 0.878 & 0.999 & 0.993 & 0.961 \\

    \sysone{XLSR+SLS~\citep{zhang2024audio}}
    & Open & 340.0 & 0.704 & 0.786 & 0.940 & 0.989 & 0.871 & 0.741 & 0.638 & 0.949 & 0.548 & 0.528 & 0.637 & 0.856 & 0.847 & 0.831 & 0.934 & 0.778 \\

    \sysone{DF\_Arena\_100M\_V\_0}
    & Closed & 100.0 & 0.700 & 0.800 & 0.970 & 0.830 & 0.640 & 0.370 & 0.600 & 0.940 & 0.840 & 0.550 & 0.820 & 0.910 & 0.860 & 0.990 & 0.990 & 0.950 \\

    \sysone{TCM~\citep{truong2024temporal}}
    & Open & 319.0 & 0.694 & 0.763 & 0.937 & 0.991 & 0.866 & 0.718 & 0.637 & 0.893 & 0.520 & 0.490 & 0.553 & 0.824 & 0.834 & 0.804 & 0.922 & 0.761 \\

    \syscite{BiCrossMamba-ST}{kheir2025bicrossmamba}
    & Open & 318.2 & 0.687 & 0.765 & 0.936 & 0.967 & 0.834 & 0.699 & 0.720 & 0.931 & 0.501 & 0.567 & 0.588 & 0.774 & 0.776 & 0.813 & 0.929 & 0.753 \\

    \sysone{Nes2NetX~\citep{liu2025nes2net}}
    & Open & 317.9 & 0.684 & 0.760 & 0.937 & 0.994 & 0.900 & 0.786 & 0.590 & 0.937 & 0.484 & 0.522 & 0.475 & 0.842 & 0.867 & 0.762 & 0.906 & 0.715 \\

    \syscite{Wav2Vec2 AASIST}{tak2022automatic}
    & Open & 317.8 & 0.651 & 0.721 & 0.909 & 0.989 & 0.960 & 0.440 & 0.677 & 0.926 & 0.443 & 0.560 & 0.623 & 0.788 & 0.840 & 0.749 & 0.693 & 0.574 \\

    \syscite{XLSR Mamba}{xiao2025xlsr}
    & Open & 319.0 & 0.637 & 0.780 & 0.946 & 0.980 & 0.955 & 0.744 & 0.708 & 0.933 & 0.528 & 0.524 & 0.577 & 0.836 & 0.854 & 0.771 & 0.926 & 0.782 \\

    \syscite{Whisper Mesonet}{kawa2023improved}
    & Open & 7.6 & 0.596 & 0.596 & 0.775 & 0.769 & 0.515 & 0.721 & 0.583 & 0.523 & 0.534 & 0.480 & 0.509 & 0.671 & 0.648 & 0.559 & 0.612 & 0.448 \\

    \syscite{Wav2Vec2 ECAPA}{kulkarni2024exploring}
    & Open & 324.0 & 0.527 & 0.460 & 0.739 & 0.328 & 0.356 & 0.161 & 0.640 & 0.377 & 0.467 & 0.398 & 0.539 & 0.724 & 0.718 & 0.118 & 0.417 & 0.388 \\

    \sysone{AASIST~\citep{jung2022aasist}}
    & Open & 0.3 & 0.483 & 0.514 & 0.625 & 0.961 & 0.607 & 0.173 & 0.425 & 0.784 & 0.368 & 0.384 & 0.397 & 0.610 & 0.754 & 0.358 & 0.318 & 0.461 \\

    \syscite{WavLM ECAPA}{kulkarni2024exploring}
    & Open & 102.0 & 0.475 & 0.573 & 0.703 & 0.964 & 0.737 & 0.227 & 0.537 & 0.767 & 0.415 & 0.420 & 0.335 & 0.778 & 0.759 & 0.490 & 0.379 & 0.615 \\

    \sysone{RawGatST~\citep{tak2021end_2}}
    & Open & 0.4 & 0.474 & 0.512 & 0.532 & 0.951 & 0.636 & 0.155 & 0.376 & 0.469 & 0.378 & 0.433 & 0.407 & 0.701 & 0.797 & 0.564 & 0.272 & 0.528 \\

    \sysone{RawTFNet~\citep{xiao2025rawtfnet}}
    & Open & 0.2 & 0.395 & 0.536 & 0.665 & 0.915 & 0.790 & 0.216 & 0.335 & 0.635 & 0.362 & 0.424 & 0.360 & 0.693 & 0.771 & 0.580 & 0.405 & 0.487 \\

    \syscite{Hubert ECAPA}{kulkarni2024exploring}
    & Open & 102.0 & 0.375 & 0.519 & 0.666 & 0.951 & 0.582 & 0.258 & 0.471 & 0.662 & 0.415 & 0.387 & 0.338 & 0.592 & 0.653 & 0.432 & 0.377 & 0.632 \\

    \sysone{Rawnet2~\citep{tak2021end}}
    & Open & 17.6 & 0.348 & 0.363 & 0.565 & 0.295 & 0.230 & 0.075 & 0.368 & 0.344 & 0.376 & 0.371 & 0.265 & 0.533 & 0.448 & 0.385 & 0.235 & 0.605 \\

    \bottomrule
    \end{tabular}
    }%
\end{table*}

\end{document}